\documentclass[fleqn,usenatbib]{mnras}

\usepackage[T1]{fontenc}

\DeclareRobustCommand{\VAN}[3]{#2}
\let\VANthebibliography\thebibliography
\def\thebibliography{\DeclareRobustCommand{\VAN}[3]{##3}\VANthebibliography}

\usepackage{graphicx}
\usepackage{amsmath}	
\usepackage{amssymb}	
\usepackage{bm}
\usepackage{hyperref}
\usepackage[normalem]{ulem}

\usepackage{newtxtext}
\usepackage[varvw]{newtxmath}

\let\vec\bm

\newcommand{\diff}{\ensuremath{\mathrm{d}}}

\def\vx{\vec{x}}
\def\vv{\vec{v}}
\def\vp{\vec{p}}

\newcommand{\ac}{a_\mathrm{c}}

\newcommand{\aeq}{a_\mathrm{eq}}

\newcommand{\ad}{a_\mathrm{d}}

\newcommand{\Td}{T_\mathrm{d}}

\newcommand{\pdc}{p_{0\mathrm{d}}}
\newcommand{\pc}{p_0}

\newcommand{\rhoM}{\rho_\mathrm{m}}

\newcommand{\fmax}{f_\mathrm{max}}
\newcommand{\rhoc}{\rho_\mathrm{c}}
\newcommand{\rc}{r_\mathrm{c}}
\newcommand{\kcut}{k_\mathrm{cut}}

\newcommand{\Lfs}{\Lambda_\mathrm{fs}}

\def\chg#1{#1}

\title[Inner cusps of the first haloes]{Inner cusps of the first dark matter haloes:\\ Formation and survival in a cosmological context}

\author[M. S. Delos \& S. D. M. White]{
M. Sten Delos,$^{1}$\thanks{E-mail: sten@mpa-garching.mpg.de}
Simon D. M. White$^{1}$
\\
$^{1}$Max Planck Institute for Astrophysics, Karl-Schwarzschild-Str. 1, 85748 Garching, Germany
}

\date{Accepted XXX. Received YYY; in original form ZZZ}

\pubyear{2021}

\begin{document}

\label{firstpage}
\pagerange{\pageref{firstpage}--\pageref{lastpage}}
\maketitle

\begin{abstract}
  We use very high resolution cosmological zoom simulations to follow the early evolution of twelve first-generation haloes formed from gaussian initial conditions with scale-free power spectra truncated on small scales by a gaussian \chg{in wavenumber}. Initial collapse occurs with a diverse range of sheet- or filament-like caustic morphologies, but in almost all cases it gives rise to a numerically converged density cusp with $\rho = Ar^{-3/2}$ and total mass comparable to that of the corresponding peak in the initial linear density field. The constant $A$ can be estimated to within about 10 per cent from the properties of this peak. This outcome agrees with earlier work on the first haloes in cold and warm dark matter universes. Within central cusps, the velocity dispersion is close to isotropic, and \chg{the} equidensity surfaces tend to align with those of the main body of the halo at larger radii. As haloes grow, their cusps are often (but not always) overlaid with additional material at intermediate radii to produce profiles more similar to the Einasto or NFW forms typical of more massive haloes. Nevertheless, to the extent that we can resolve them, cusps survive at the smallest radii. Major mergers can disturb them, but the effect is quite weak in the cases that we study. The cusps extend down to the resolution limits of our simulations, which are typically a factor of several larger than the cores that would be produced by phase-space conservation if the initial power spectrum cutoff arises from free streaming.
\end{abstract}




\section{Introduction}

In the idealized cold dark matter (CDM) picture, density variations are present at all scales within the initial mass distribution. These density variations seed gravitationally collapsed haloes that form hierarchically through accretion and mergers of smaller haloes. Numerical simulations of this scenario have found that the spherically averaged density profiles $\rho(r)$ of these haloes are remarkably universal as a function of the radius $r$. \citet{1996ApJ...462..563N,1997ApJ...490..493N} showed that all CDM haloes are well described by a functional form -- now called the Navarro--Frenk--White (NFW) profile -- that approaches $\rho\propto r^{-1}$ at small radii and steepens towards $\rho\propto r^{-3}$ as the outer "virial" radius of the halo is approached. \citet{2004MNRAS.349.1039N} later found that the Einasto density profile \citep{1965TrAlm...5...87E}, in which the logarithmic density slope varies with radius as $\diff\log\rho/\diff\log r\propto -r^{\alpha_\mathrm{E}}$ with $\alpha_\mathrm{E}\sim 0.17$, supplies a slightly better fit. This profile's density slope becomes even shallower at small radii than the NFW asymptotic limit of $-1$ \citep{2010MNRAS.402...21N}.

This picture is incomplete, however. Most models of particle dark matter possess some thermal motion in the initial conditions. This motion smooths out the primordial density field on a characteristic, free-streaming scale. For example, if the dark matter is the lightest neutralino in a supersymmetric extension of the Standard Model with a mass of order 100~GeV, then the free-streaming scale corresponds to roughly an Earth mass, and the first haloes form at about this mass \citep[e.g.][]{2005JCAP...08..003G}. For warm dark matter (WDM) models, the free-streaming mass scale could be as large as  $\sim10^8~\mathrm{M}_\odot$ \citep[e.g.][]{2021MNRAS.506.5848E}. The first haloes form via the direct monolithic collapse of smooth density peaks at this scale. Compared to the NFW or Einasto density profiles associated with idealized CDM haloes, the density profiles of these first haloes tend to be steeper at small radii, with slopes approaching $\diff\log\rho/\diff\log r = -3/2$ \citep{2005Natur.433..389D,2010ApJ...723L.195I,2013JCAP...04..009A,2014ApJ...788...27I,2015MNRAS.450.2172P,2017MNRAS.471.4687A,2018MNRAS.473.4339O,2018PhRvD..97d1303D,2018PhRvD..98f3527D,2020MNRAS.492.3662I,2021A&A...647A..66C}.\footnote{Dark matter halo density profiles that approach $\rho\propto r^{-3/2}$ at small radii were suggested earlier by \citet{1999MNRAS.310.1147M} in the context of idealized CDM, but later works \citep[e.g.][]{2004MNRAS.349.1039N} found that this behaviour did not hold as simulation resolution improved.}

Yet the relevance of these $\rho\propto r^{-3/2}$ cusps to cosmological haloes remains uncertain. The central density slope tends to be shallower for more massive haloes \citep{2014ApJ...788...27I,2017MNRAS.471.4687A}, which suggests that steep inner cusps may relax into shallower NFW or Einasto forms over time as haloes grow. Moreover, a separate body of work \citep{1999MNRAS.308.1011H,2001ApJ...559..516A,2007ApJ...665....1B,2009MNRAS.396..709W,2014MNRAS.439..300L,2020Natur.585...39W} continues to find no evidence of steep central cusps in even the smallest haloes, instead concluding that these haloes possess the same NFW or Einasto density profiles as their idealized CDM counterparts. It is also noteworthy that simulations that resolve the free-streaming scale are polluted by smaller-scale structures that arise artificially from discreteness noise \citep[e.g.][]{2007MNRAS.380...93W}, an issue that could cast doubt on the numerical robustness of the density profiles in these simulations.

The present study is aimed at clarifying the extent to which $\rho\propto r^{-3/2}$ central density cusps arise and persist in a cosmological context. We carry out very high resolution zoom simulations of twelve first-generation haloes, grown from three different initial linear power spectra, and we track the structure and the density profiles of these haloes through their formation and early evolution. The details of the initial collapse differ dramatically in the twelve cases, as does the degree of artificial structuring present during the formation process. Our first objective is to test whether these diverse formation sequences all result in haloes with $\rho\propto r^{-3/2}$ in their inner regions.

Our second goal is to explore the survival of these cusps. Appealing to the results of idealized simulations, \citet{2016MNRAS.461.3385O} and~\citet{2017MNRAS.471.4687A} argued that major mergers drive the transition of $\rho\propto r^{-3/2}$ cusps toward the shallower NFW or Einasto forms found in larger haloes. On the other hand, studies of CDM haloes by \citet{2010arXiv1010.2539D} and \citet{2013MNRAS.432.1103L}, among others, found the overall mass accretion history of a halo, which is usually dominated by minor mergers and diffuse accretion rather than by major mergers  \citep[e.g.][]{2010MNRAS.401.1796A}, to be a good predictor of its density profile. Our halo sample exhibits a variety of accretion and merger histories, the range of which is enhanced by our sampling of very different initial power spectra. We explore how the later-time density profiles of our haloes vary with accretion and merger history, and we investigate how this affects their initial central cusps.

If they survive, $\rho\propto r^{-3/2}$ central cusps might have significant observational implications. Subhaloes with steeper central cusps are substantially more resistant to tidal disruption \citep{2010MNRAS.406.1290P,2022arXiv220700604S}, so $\rho\propto r^{-3/2}$ cusps could boost the abundance of the lowest-mass subhaloes in the haloes of the Milky Way and other large galaxies. Steeper cusps are also associated directly with stronger observational signatures. For instance, prospects for the detection of dark subhaloes using
annihilation radiation are enhanced because the annihilation luminosity diverges logarithmically towards $r=0$ for a profile with $\rho\propto r^{-3/2}$ \chg{(for velocity-independent cross sections)}. This behaviour illustrates the importance of understanding the range of radii encompassed by the cusps of the first haloes. Our final aim is thus to explore this question. Our simulations can determine the maximum radii of the cusps, but they do not resolve the minimum radius. However, if the initial density field was smoothed by free streaming of the dark matter particles, then these same motions must tame the central cusp by imposing a core at the maximum phase-space density allowed by Liouville's theorem \citep{1979PhRvL..42..407T}. We study the phase-space structure of the inner cusp analytically to determine the radius at which it must give way to a core.

Our main results support the robustness of  $\rho\propto r^{-3/2}$ cusps in the first haloes. We find that such cusps arise for haloes with wildly different formation sequences and are well converged with respect to simulation resolution. As our haloes grow, rapid accretion is the main effect leading to shallower density profiles, but this process works by building up the density at intermediate radii without substantially altering the inner cusp. Consequently, haloes can develop apparently shallow Einasto-like density profiles at large and intermediate radii while retaining the initial $\rho\propto r^{-3/2}$ cusp at the smallest radii. Major merger events can disturb this central cusp, but their impact is relatively minor in the cases that we have studied. Finally, the radius at which phase-space constraints force a cusp to give way to a core is sufficiently small in realistic models that the $\rho\propto r^{-3/2}$ cusp can span several decades in radius. These findings suggest that $\rho\propto r^{-3/2}$ central cusps may survive long enough to be observationally relevant, although studies involving larger halo samples and longer time spans (and hence larger growth factors) are needed to be certain.

This article is organized as follows. In Section~\ref{sec:setup}, we discuss our simulation setup and the properties of our simulated haloes. In Section~\ref{sec:form}, we explore halo formation mechanisms and the density cusps that emerge. In Section~\ref{sec:evo}, we explore how and why halo density profiles evolve over time, and we study the extent to which the $\rho\propto r^{-3/2}$ cusps survive this process. In Section~\ref{sec:core}, we use analytic arguments to evaluate the radius at which an inner density cusp must give way to a core with finite central density. We summarize our results in Section~\ref{sec:conclusion}. Finally, Appendices \ref{sec:lowres} and~\ref{sec:convergence} validate the numerical robustness of our simulation results, while Appendix~\ref{sec:structure} presents mathematical details relating to the calculation of the core size in Section~\ref{sec:core}.

\section{Simulation setup}\label{sec:setup}

\subsection{Power spectra and units}

\begin{figure}
	\centering
	\includegraphics[width=\columnwidth]{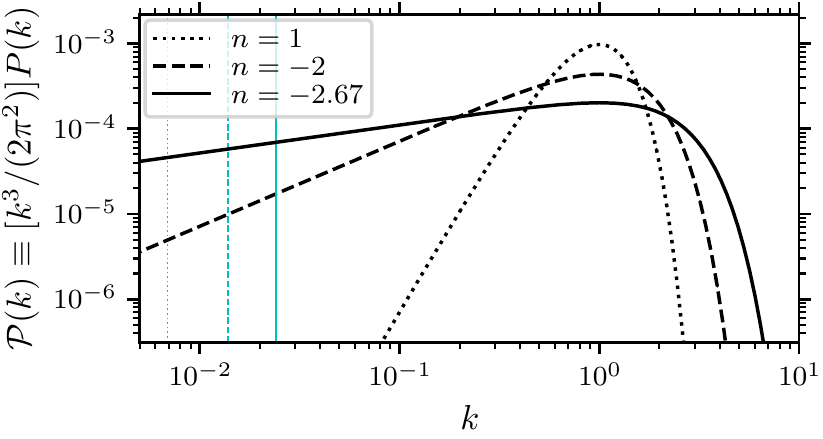}
	\caption{The power spectra from which we draw our simulation initial conditions. We show the dimensionless form $\mathcal{P}(k)\equiv [k^3/(2\pi^2)]P(k)$. The length units are fixed such that $\langle k^2\rangle=1$, so the gaussian cutoff scale is $\kcut=\sqrt{2/(3+n)}$. These spectra are normalized so that the rms density contrast is $\sigma=0.03$. The vertical cyan lines indicate the fundamental frequency associated with the box employed for simulations with each power spectrum.}
	\label{fig:pk}
\end{figure}

To explore scenarios in which density variations are cut off at small scale, we consider Einstein-de Sitter cosmologies with initial power spectra taking the form
\begin{align}\label{pk}
    P(k) \propto k^n \exp[-(k/\kcut)^2],
\end{align}
i.e. power laws multiplied by a gaussian of scale  $k=\kcut$.\footnote{A cutoff of exactly this form arises from free streaming if the thermal velocity distribution is Maxwell-Boltzmann \citep[e.g.][]{2006PhRvD..74f3509B}.} Specifically, we consider the three power-law indices $n=1$, $n=-2$, and $n=-2.67$. These power spectra are depicted in Fig~\ref{fig:pk}; we show the dimensionless form $\mathcal{P}(k)\equiv [k^3/(2\pi^2)]P(k)$, indicating the typical (squared) relative amplitude of density variations as a function of their scale. Generally, power spectra with $n$ close to $-3$ are associated with the presence of significant density variations over a broad range of scales. In contrast, the $n=1$ power spectrum has significant support only within a narrow range. In standard cosmological scenarios, cold dark matter is associated with $n\sim -3$ at scales close to the cutoff while warmer dark matter is associated with $n$ farther from $-3$. For instance, we will see later that the power spectrum associated with $3.5$-keV warm dark matter \citep[e.g.][]{2001ApJ...556...93B} lies somewhere between the $n=-2.67$ and $n=-2$ power laws near the cutoff. Substantially larger power-law indices are not observationally viable in this context, although they arose in the original hot dark matter models \citep[e.g.][]{1983ApJ...274L...1W} and are also found for cold dark matter for some nonstandard cosmological initial condition scenarios \citep[e.g.][]{2011PhRvD..84h3503E}. In this paper, we do not aim to reproduce any particular cosmology but rather to explore the formation of first-generation objects using idealized initial power spectra.

The power spectra in Fig.~\ref{fig:pk} are normalized so that the (unfiltered) rms density contrast is $\sigma=0.03$, where
\begin{equation}
    \sigma^2=\int_0^\infty\frac{\diff k}{k}\mathcal{P}(k).
\end{equation}
We also fix our length units so that $\langle k^2\rangle=1$, where
\begin{equation}\label{k2}
    \langle k^2\rangle = \sigma^{-2} \int_0^\infty\frac{\diff k}{k}k^2\mathcal{P}(k),
\end{equation}
which leads to $\kcut=\sqrt{2/(3+n)}$ and puts the peak of $\mathcal{P}$ at $k=1$. The choice to fix $\sigma$ and $\langle k^2\rangle$ is
\chg{convenient because it means that density peaks arising from different power spectra have comparable amplitudes and sizes. Within the linear density field $\delta(\vx)$, peak amplitudes $\delta$ are proportional to $\sigma$, while peak sizes, characterized by $|\delta/\nabla^2\delta|^{1/2}$, cluster around $\langle k^2\rangle^{-1/2}$ in the high-peak limit \citep{1986ApJ...304...15B}. As we will see in Section~\ref{sec:peaks}, the inner structures of the first haloes are sensitive almost entirely to these quantities.}\footnote{\chg{If we were interested in halo \emph{counts} instead of structures, the more natural length scale would instead be $\langle k^4\rangle^{-1/2}\langle k^2\rangle^{1/2}=\sqrt{2/(5+n)}\kcut^{-1}$, which sets the number density of peaks \citep{1986ApJ...304...15B}. On the other hand, the ``half-mode'' scale commonly considered in studies of warm dark matter \citep[e.g.][]{2021MNRAS.506.5848E}, where the cutoff suppresses the amplitudes of density variations by half, has no clear meaning for haloes. The half-mode mass in our units is $M_\mathrm{hm}=1800$, $M_\mathrm{hm}=225$, and $M_\mathrm{hm}=42.7$ for the Eq.~(\ref{pk}) power spectrum with $n=1$, $n=-2$, and $n=-2.67$, respectively; compare to Fig.~\ref{fig:halomass}.}}

\begin{table}
	\centering
	\caption{The units that we employ in this study. We specify comoving units here; for physical units the scale factor $a$ is defined such that $a=\sigma$ in linear theory.}
	\label{tab:units}
	\begin{tabular}{cc}
		\hline
		quantity & unit \\
		\hline
		length & $\langle k^2\rangle^{-1/2}$ \\
		density & $\bar\rho$ \\
		mass & $\bar\rho\langle k^2\rangle^{-3/2}$ \\
		\hline
	\end{tabular}
\end{table}

In our simulations we will take the initial scale factor to be $a=0.03$, so that $a=\sigma$ in linear theory. We also define mass units such that the mean comoving cosmological density is 1. The units and conventions are summarized in Table~\ref{tab:units}.

\subsection{Simulation procedure}

\begin{figure}
	\centering
	\includegraphics[width=\columnwidth]{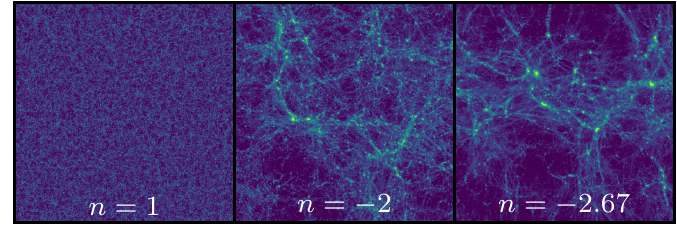}
	\caption{Projected density field within a representative region of each simulation at $a=2$ (when the {\it rms} density contrast $\sigma=2$ in linear theory). The region is a cube of width $261\langle k^2\rangle^{-1/2}$; for $n=-2.67$ this is the entire box. The vastly higher amplitude of large-scale power in the $n=-2$ and $n=-2.67$ simulations, compared to the $n=1$ simulation, is immediately evident. The latter produces an essentially uniform distribution of haloes, while the former produce structures that approach the size of the displayed region.}
	\label{fig:fields}
\end{figure}

We use the \textsc{Gadget-4} simulation code \citep{2021MNRAS.506.2871S} to execute our simulations. First, we carry out one primary simulation for each of the three power spectra in Fig.~\ref{fig:pk}. In each case we use second-order Lagrangian perturbation theory to produce a perturbed grid of $1024^3$ particles inside a periodic box of width $643\kcut^{-1}$. This choice means that the Nyquist frequency associated with the initial particle grid is $5\kcut$. We then simulate each box up to some scale factor $a>2$ under the assumption of matter domination. The simulation force softening length here is set to 0.03 times the initial interparticle spacing. A portion of each resulting simulation box at $a=2$ is shown in Fig.~\ref{fig:fields}; in particular we show a cube of width $261\langle k^2\rangle^{-1/2}$, which for $n=-2.67$ is the entire box. The imprint of the power spectrum is evident here. The $n=1$ simulation's dearth of large-scale power yields an essentially uniform distribution of tiny haloes, whereas the substantial amplitude of large-scale power in the $n\leq -2$ simulations produces much larger haloes and filaments; in the $n=-2.67$ case these structures begin to approach the box size.

From each of these primary simulations, we select a range of haloes to resimulate at much higher resolution. The selection is mostly arbitrary, but we favor haloes that formed by $a\sim 0.6$. We aim to ensure that there is a reasonable sense in which each halo initially formed by direct collapse. To boost the resolution, we identify each low-resolution particle that resides within the halo's spherical-overdensity radius $R_{178}$ (enclosing an average density 178 times the cosmological mean) at some late time. We then locate that particle within the initial conditions and replace it with a glass consisting of between $4.0\times 10^3$ and $2.6\times 10^5$ high-resolution particles (depending on the halo), \chg{whose center of mass lies at the original particle's location. We also remove low-resolution ``holes'' in the high-resolution patch by iteratively ensuring that for any particle selected to be replaced by a glass, we also select its nearest neighbor along the cardinal direction that most closely points toward the high-resolution region's center of mass.}

For each such high-resolution halo, we then carry out another simulation that includes that halo's high-resolution particles along with the remaining original low-resolution particles (which retain their original mass and force-softening length). Note that due to the short duration of our simulations and our focus on inner density profiles, it is not necessary to ensure that low-resolution particles never enter the high-resolution halo. We discuss in Appendix~\ref{sec:lowres} the extent to which low-resolution particles do enter our haloes, showing that they remain at radii too large to affect our conclusions.

\begin{table}
	\centering
	\caption{Numerical parameters associated with our high-resolution halo simulations. For each halo (and associated power spectral index $n$), we list the number $N_\mathrm{high}$ of high-resolution particles, their mass $m_p$, the softening length $\epsilon$, and the final expansion factor and halo mass, $a_\mathrm{f}$ and $M_\mathrm{f}$.}
	\label{tab:halos}
	\tabcolsep=0.16cm
	\begin{tabular}{lcccccc}
\hline
$\hphantom{-}n$ & halo & $N_\mathrm{high}$ & $m_p$ & $\epsilon$ & $a_\mathrm{f}$ & $M_\mathrm{f}$ \\
\hline
$\hphantom{-}1$ & H1 & $2.4 \times 10^{8}$ & $2.1 \times 10^{-5}$ & $4.3 \times 10^{-4}$ & $4.72$ & $1.4 \times 10^{3}$\\
$\hphantom{-}1$ & H2 & $1.5 \times 10^{8}$ & $2.1 \times 10^{-5}$ & $4.3 \times 10^{-4}$ & $5.06$ & $2.1 \times 10^{3}$\\
$\hphantom{-}1$ & H3 & $5.6 \times 10^{7}$ & $2.1 \times 10^{-5}$ & $4.3 \times 10^{-4}$ & $6.51$ & $9.5 \times 10^{2}$\\
$-2$ & W1 & $3.0 \times 10^{8}$ & $2.1 \times 10^{-5}$ & $4.3 \times 10^{-4}$ & $1.50$ & $3.1 \times 10^{3}$\\
$-2$ & W2 & $1.9 \times 10^{8}$ & $2.1 \times 10^{-5}$ & $4.3 \times 10^{-4}$ & $1.28$ & $1.8 \times 10^{3}$\\
$-2$ & W3 & $1.4 \times 10^{8}$ & $2.1 \times 10^{-5}$ & $4.3 \times 10^{-4}$ & $1.69$ & $9.8 \times 10^{2}$\\
$-2$ & W4 & $1.6 \times 10^{8}$ & $2.1 \times 10^{-5}$ & $4.3 \times 10^{-4}$ & $1.68$ & $1.4 \times 10^{3}$\\
$-2$ & W5 & $5.9 \times 10^{8}$ & $3.3 \times 10^{-7}$ & $1.1 \times 10^{-4}$ & $1.52$ & $1.0 \times 10^{2}$\\
$-2.67$ & C1 & $8.6 \times 10^{8}$ & $4.1 \times 10^{-6}$ & $2.5 \times 10^{-4}$ & $1.00$ & $2.3 \times 10^{3}$\\
$-2.67$ & C2 & $5.6 \times 10^{8}$ & $4.1 \times 10^{-6}$ & $2.5 \times 10^{-4}$ & $1.19$ & $2.0 \times 10^{3}$\\
$-2.67$ & C3 & $2.2 \times 10^{8}$ & $4.1 \times 10^{-6}$ & $2.5 \times 10^{-4}$ & $1.03$ & $8.4 \times 10^{2}$\\
$-2.67$ & C4 & $4.7 \times 10^{8}$ & $5.1 \times 10^{-7}$ & $1.2 \times 10^{-4}$ & $1.41$ & $1.6 \times 10^{2}$\\
\hline
	\end{tabular}
\end{table}

\begin{figure}
	\centering
	\includegraphics[width=\columnwidth]{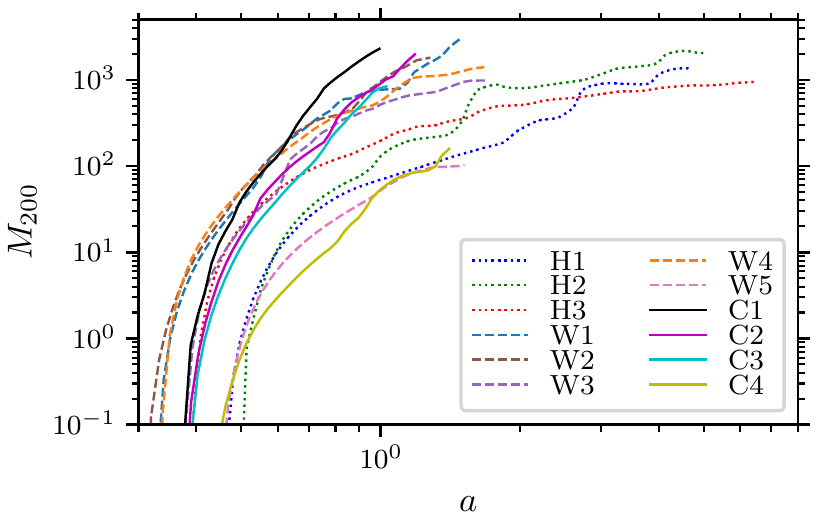}
	\caption{The spherical-overdensity mass $M_{200}$ (within a spherical region enclosing a mean density 200 times the cosmic average) of each halo that we study, as a function of the scale factor $a$. The curve associated with each halo terminates where we end the simulation.
	}
	\label{fig:halomass}
\end{figure}

The numerical parameters associated with the high-resolution halo simulations are listed in Table~\ref{tab:halos}. \chg{Since the force-softening length represents a major resolution bottleneck (see Appendix~\ref{sec:convergence}), we set the softening length of the high-resolution particles to be only about $0.016$ their initial spacing.} We simulate twelve haloes in this way, which we name H1--H3 (from the $n=1$ power spectrum), W1--W5 (from the $n=-2$ power spectrum), and C1--C4 (from the $n=-2.67$ power spectrum). The mass accretion histories of these haloes are depicted in Fig.~\ref{fig:halomass} over the period that we simulate them at high resolution.
\chg{Note that this figure already demonstrates the value of our choice of units; by fixing the characteristic sizes $|\delta/\nabla^2\delta|^{1/2}\sim 1$ of peaks in the linear density field, we ensure that all haloes have masses of order unity at the time they collapse.}

In later sections, we plot halo density profiles down to a minimum radius that is determined by the simulation particle count and the force softening length $\epsilon$. Specifically, we consider the two radii $r_\mathrm{rel}$, at which the age of the universe is equal to the two-body relaxation time scale $t_\mathrm{rel}$ (Eq.~\ref{trel}), and $r_\mathrm{soft} = 5\epsilon$. We then plot density profiles only at radii larger than both $r_\mathrm{rel}$ and $r_\mathrm{soft}$. We test a density profile for numerical convergence in Appendix~\ref{sec:convergence} and find that these choices determine reasonably well the minimum radius that is converged with respect to simulation resolution.

\section{Formation of the inner cusp}\label{sec:form}

\begin{figure}
	\centering
	\includegraphics[width=\columnwidth]{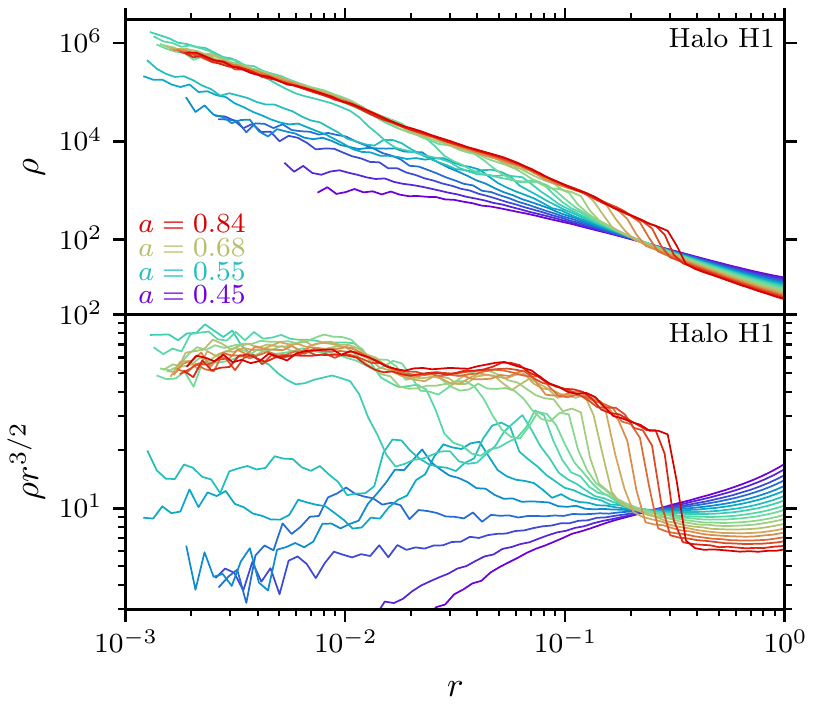}
	\caption{Density profile of halo H1 during its collapse. The upper panel shows density $\rho$ as a function of radius $r$, while the lower panel scales the vertical axis by $r^{3/2}$. We plot physical, not comoving, quantities. Different colours represent different times. The density profile begins to stabilize around $a=0.55$, after which it is essentially fixed and only grows outward in radius. The power law of the inner profile is about $\rho\propto r^{-3/2}$ (horizontal on the lower panel).}
	\label{fig:H1coll}
\end{figure}

\begin{figure*}
	\centering
	\includegraphics[width=\linewidth]{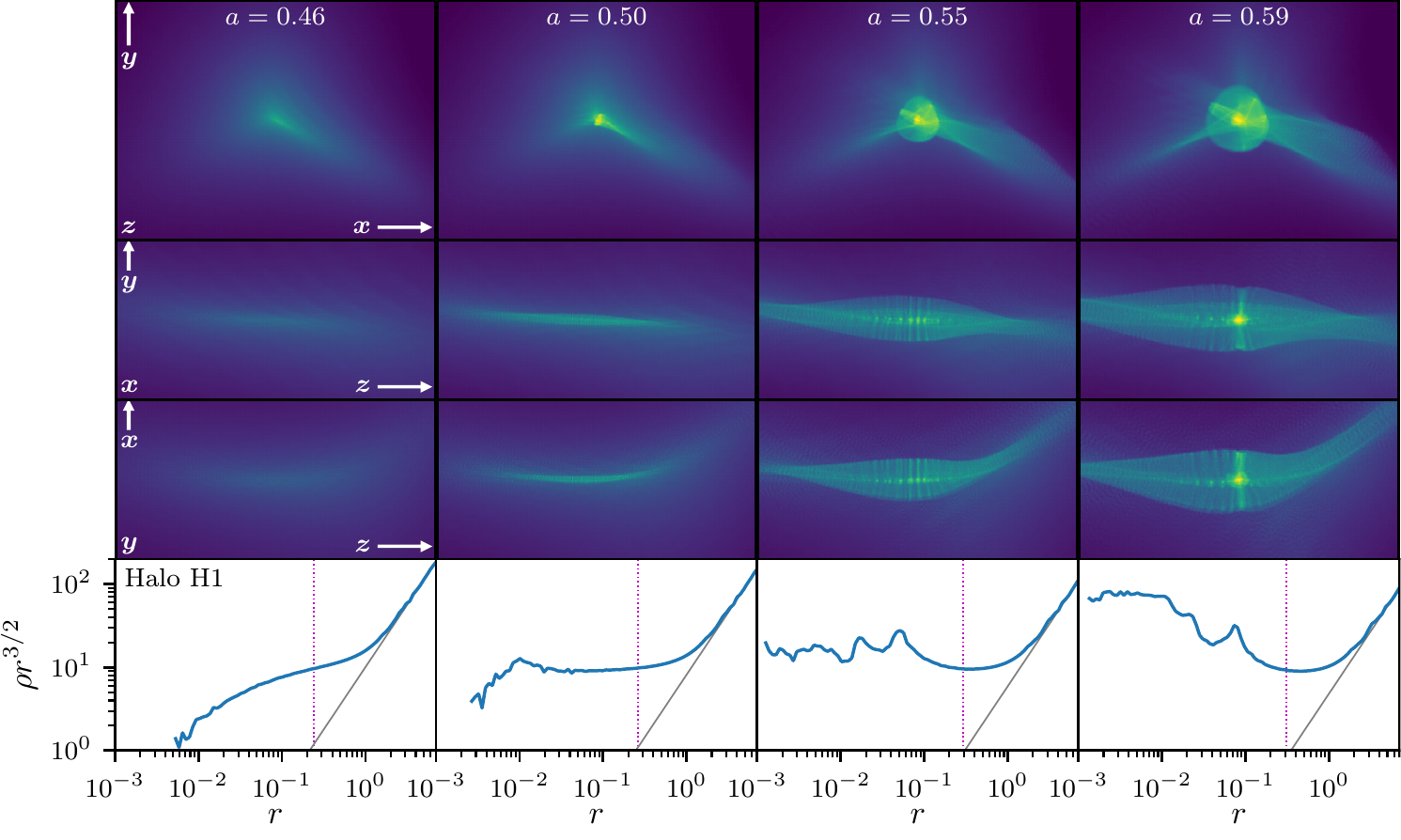}
	\caption{Formation of the halo H1. We plot three different orthogonal projections of the density field (upper three rows) at four different times (columns). These projections are aligned with the principal axes of the initial tidal tensor. Material initially collapses to form a filament (first three columns), which is evident because two projections show a linearly extended object while the third suggests a point-like object. Within this filament the collapsed halo eventually forms (last column). Fragmentation of the filament visible in the third column is a numerical artefact. In the bottom row, we show the spherically averaged density profile corresponding to the pictures above. Evidently, by $a=0.59$ (last column) the halo's stable $\rho\simeq 60 r^{-3/2}$ density profile is already present. Within the lower panels, the vertical line marks half the width of the displayed box (i.e. the radius from the centre to the left or right edge), while the faint grey diagonal line indicates the cosmological mean density. In the upper panels of the left column we include directional references to clarify the relationship between the three projections; for instance the upper panels are projected in the ``$z$'' direction, and the ``$x$'' and ''$y$'' directions therein are indicated by arrows.}
	\label{fig:H1field}
\end{figure*}

\begin{figure*}
	\centering
	\includegraphics[width=\linewidth]{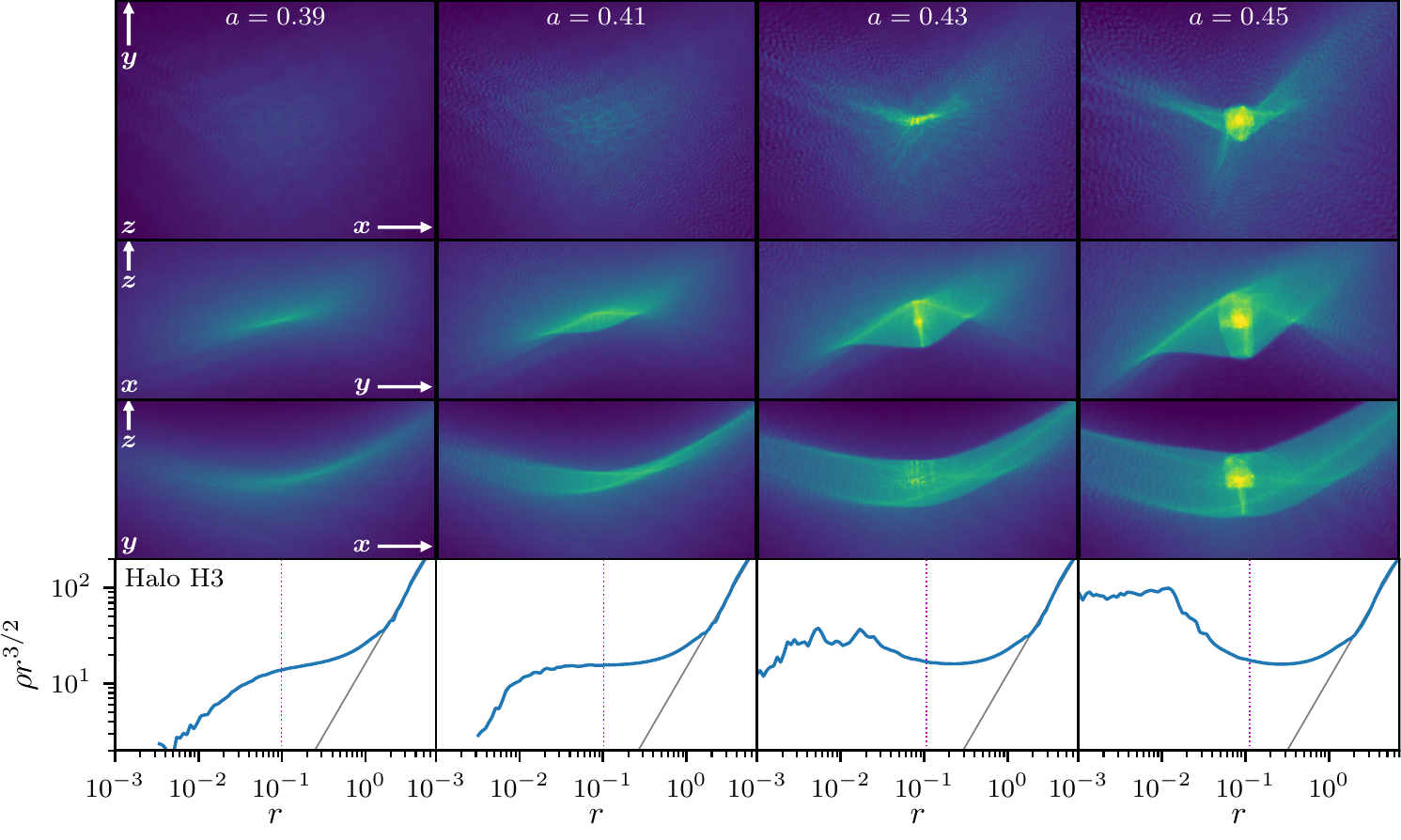}
	\caption{Formation of the halo H3. As in Fig.~\ref{fig:H1field}, we plot the density field projected along the principal axes of the initial tidal tensor (upper three panels) along with the spherically averaged density profile (lower panels) at four different times (columns); see that figure's caption for further explanation. The material in this case collapses into a sheet (first two columns), which is evidenced by a linearly extended object visible along two of the principal axes and no collapsed object visible along the third. Artificial small-scale structure is visible in the sheet in the top image of the second and third columns and in the lower image (``$y$'' projection) of the third column, but it is far less pronounced than the filament fragmentation seen in Fig.~\ref{fig:H1field}. The halo's stable inner $\rho\simeq 90 r^{-3/2}$ density profile has already developed by $a=0.45$ (last column).}
	\label{fig:H3field}
\end{figure*}

\begin{figure*}
	\centering
	\includegraphics[width=\linewidth]{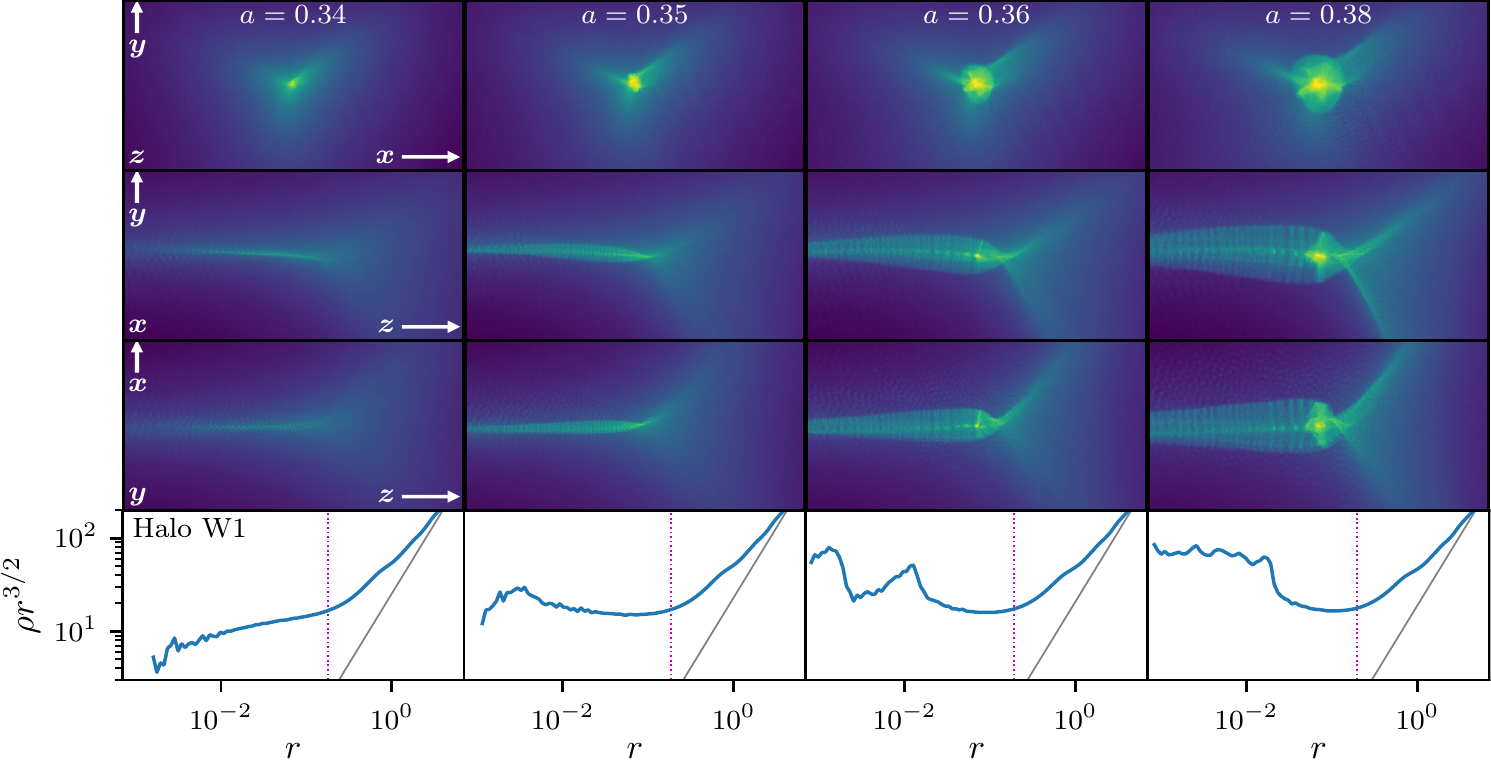}
	\caption{Formation of the halo W1. We plot the density field projected along the principal axes of the initial tidal tensor (upper three rows of panels) along with the spherically averaged density profile (lower panels) at four different times (columns); see the caption of Fig.~\ref{fig:H1field} for further explanation. The halo's material initially collapses into a filament (first two columns), similarly to the case of the H1 halo (Fig.~\ref{fig:H1field}). Unlike H1, however, W1 forms at the end of its filament (third column). Moreover, while artificial fragmentation is visible at this time, the halo has clearly already formed, as evidenced by comparing the density profiles in the third and fourth columns at the innermost resolved radii. By $a=0.38$ (last column) a $\rho\simeq 70 r^{-3/2}$ density profile is already well developed.}
	\label{fig:W1field}
\end{figure*}

\begin{figure*}
	\centering
	\includegraphics[width=\linewidth]{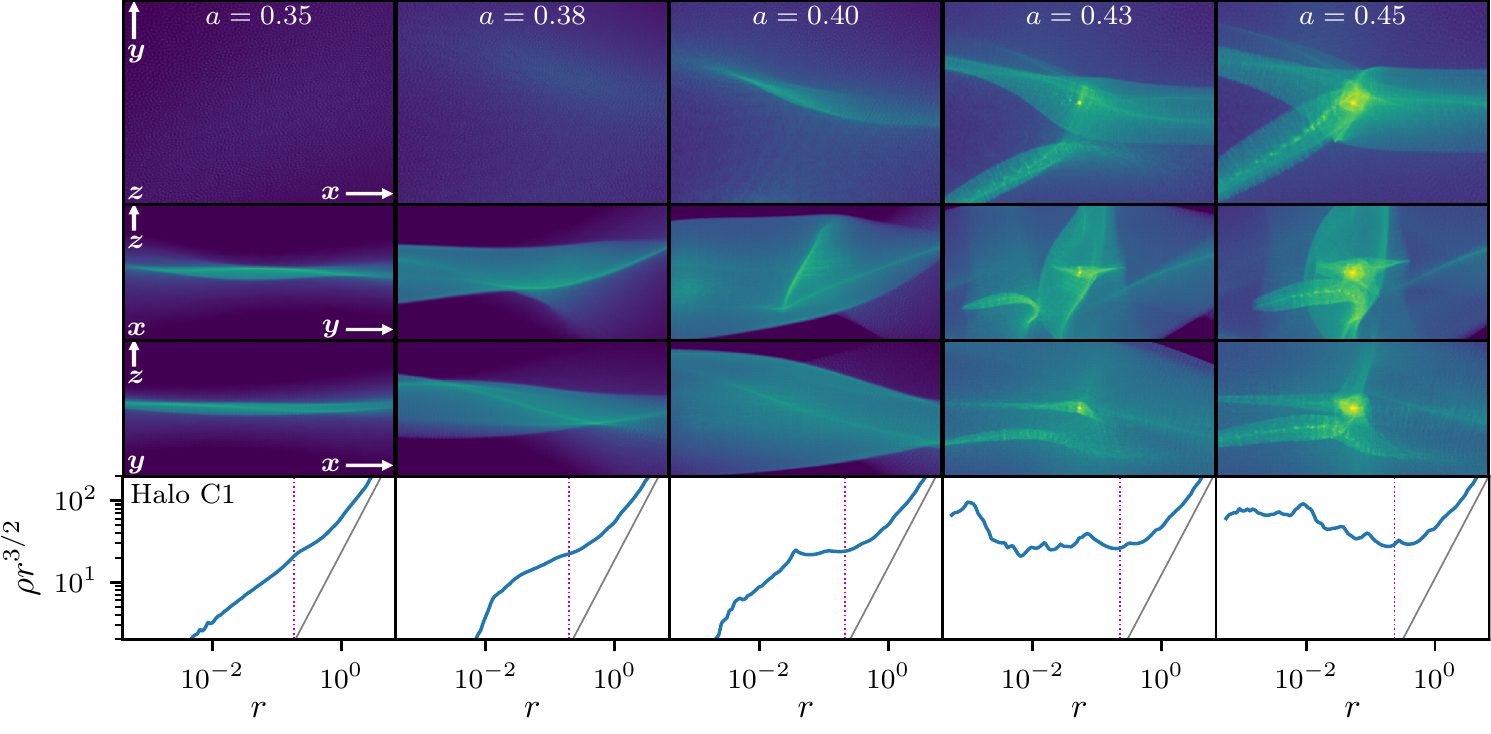}
	\caption{Formation of the halo C1. We plot the density field projected along the principal axes of the initial tidal tensor (upper three rows of panels) along with the spherically averaged density profile (lower panels) at five different times (columns); see the caption of Fig.~\ref{fig:H1field} for further explanation. This halo's material initially collapses into a sheet (first column), which is far more pronounced than the sheet associated with the H3 halo (Fig.~\ref{fig:H3field}). This sheet soon collapses into a filament (third column) followed almost immediately by the halo (fourth column). The halo's $\rho\simeq 70 r^{-3/2}$ density profile is well developed by $a=0.45$ (last column); meanwhile, another filament is about to accrete edge-on, although its accretion will not change the density profile significantly (see Fig.~\ref{fig:profiles-coll}).}
	\label{fig:C1field}
\end{figure*}

\begin{figure*}
	\centering
	\includegraphics[width=\linewidth]{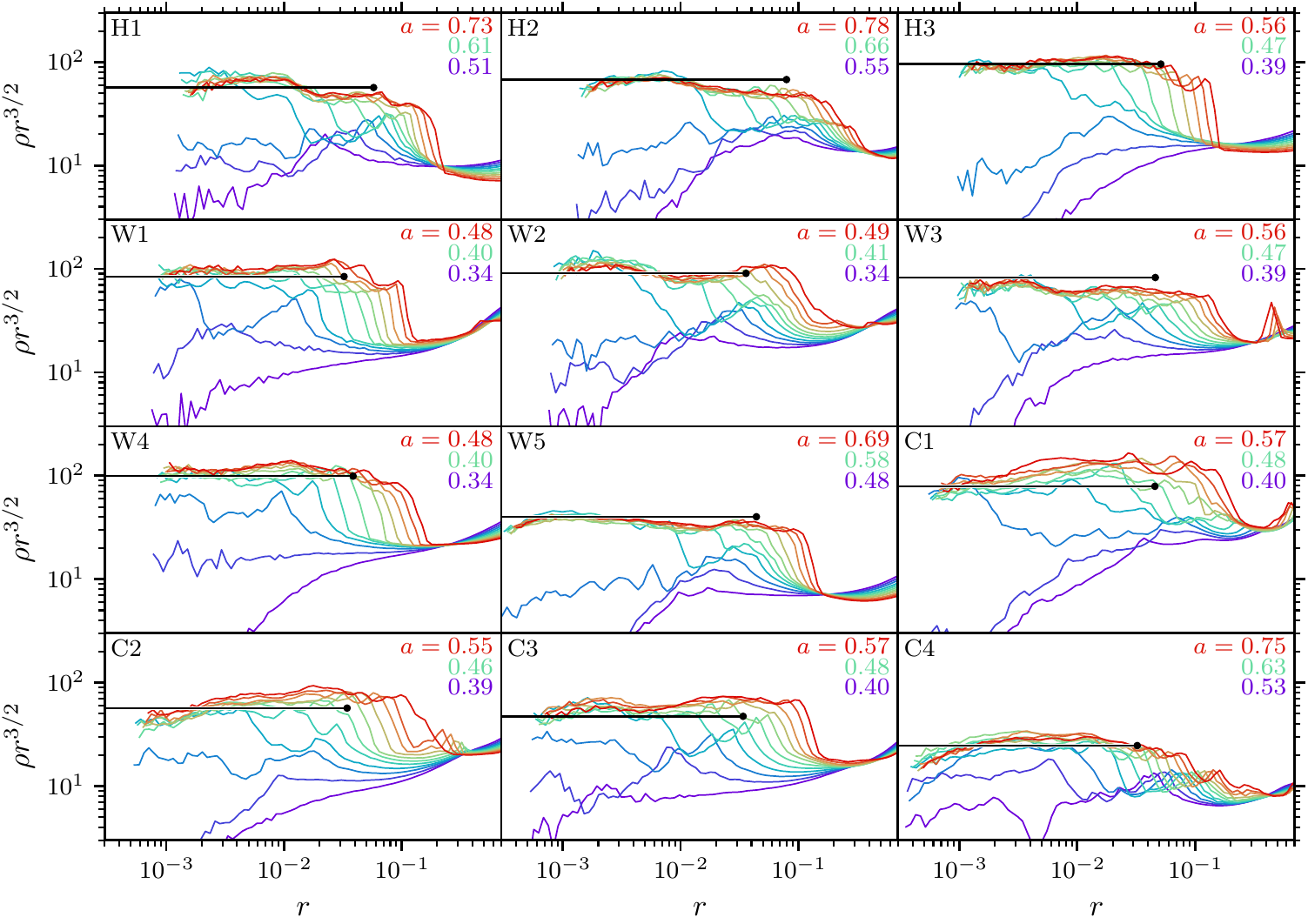}
	\caption{Density profiles of all of our high-resolution haloes during their collapse. As in Fig.~\ref{fig:H1coll}, the vertical axis is scaled by $r^{3/2}$; we plot physical, not comoving, quantities; and different colours represent different times. The haloes all develop $\rho\propto r^{-3/2}$ density cusps very quickly after they collapse. In some cases there is a tendency for the profile to shallow at the smallest radii, but it is unclear whether this behaviour is numerically converged (see Appendix~\ref{sec:convergence}). The horizontal black line indicates the cusp amplitude $A=\rho r^{3/2}$ predicted from each halo's precursor density peak using Eq.~(\ref{A}) with $\alpha=24$. This line terminates at the estimated initial cusp radius $r_0$ predicted from the same peak data using Eq.~(\ref{r0}). There is some scatter, but overall these predictions match well the actual properties of the cusps.}
	\label{fig:profiles-coll}
\end{figure*}

We now explore the formation of the central cusp during the initial collapse of the high-resolution haloes presented above. A major goal of this section is to show that a wide variety of collapse sequences all give rise to the same power-law $\rho\propto r^{-3/2}$ inner density profile. We also discuss the connection between this feature and the initial density field.

\subsection{The formation process}

We first focus on haloes arising from the $n=1$ power spectrum. Figure~\ref{fig:H1coll} shows the density profile of the H1 halo during and shortly after its formation. This halo forms around $a\simeq 0.6$, and its inner density profile immediately settles at $\rho\simeq 60 r^{-3/2}$ (in physical coordinates). From this point on, the density profile essentially remains stable at small radii, only growing outward due to new accretion. This picture closely resembles that in \citet{2019PhRvD.100b3523D}, even with the present study's significantly improved spatial resolution: \chg{for $a\gtrsim 0.65$, there are $3\times 10^5$ simulation particles at the radii $r<0.06$ corresponding to the $\rho\propto r^{-3/2}$ part of the density profile.}

Figure~\ref{fig:H1field} shows the halo's surrounding density field during its formation. We show three orthogonal projections at each time, and to clarify the picture, we align these projections with the principal axes of the tidal tensor at the halo's Lagrangian position within the initial density field. Evidently, the halo's precursor overdensity first collapses into a one-dimensional filament (second column in Fig.~\ref{fig:H1field}) aligned in the  $z$ direction with very little extent in either the $x$ or the $y$ direction. This filament begins to break up into regular fragments, easily visible in the third column, which are a well-known numerical artefact in simulations of this type \citep[e.g.][]{2007MNRAS.380...93W,2013MNRAS.434.3337A,2014MNRAS.439..300L}. Finally, a collapsed halo forms near the densest point of the filament. For convenience we also plot the spherically averaged density profile at each time. By $a=0.59$ the inner $\rho\simeq 60 r^{-3/2}$ cusp is already present. The halo's outer density caustic is clearly visible in the profile  at this time around $r\simeq 2\times 10^{-2}$, and the filament's outer caustic can also be seen around $r\simeq 8\times 10^{-2}$ (the latter will soon be swallowed by the growing halo; see Fig.~\ref{fig:H1coll}).

In Fig.~\ref{fig:H1field}, the inner cusp could be interpreted as forming from a collection of artificial filament fragments, a prospect that might cast doubt on the validity of the central profile of the resulting halo. These fragments arise because the discrete simulation particles that comprise the filament tend to cluster (a process that may be viewed as the triggering of a one-dimensional Jeans instability by discreteness noise).\footnote{Simulation methods exist that can eliminate these artefacts \citep{2020MNRAS.495.4943S,2022MNRAS.509.1703S}, although we do not employ them here.} However, two lines of evidence suggest that the $\rho \propto r^{-3/2}$ cusp is robust.
\begin{enumerate}
    \item In Appendix~\ref{sec:convergence}, we explicitly test the density profile of the W5 halo for numerical convergence. This halo also exhibits a clear $\rho\propto r^{-3/2}$ power-law density profile at small radii, and its profile is stable with respect to changes in the simulation's spatial and mass resolution. In contrast, the mass and number of artificial filament fragments scale with a simulation's linear resolution \citep{2007MNRAS.380...93W}.
    \item As we will see next, a halo that collapses from a two-dimensional sheet (instead of a one-dimensional filament) also develops a clear $\rho\propto r^{-3/2}$ density cusp, even though the sheet exhibits far weaker artificial structuring.
\end{enumerate}

Figure~\ref{fig:H3field} depicts the density field around the H3 halo during its collapse. We show again the field projected along the tidal tensor's principal axes. The material in this case collapses into a sheet (first two columns), as indicated by two of the  projections showing a linearly extended object while the third shows no apparent collapsed object at all.\footnote{The principal-axis projections are particularly valuable in the case of sheet collapse, as most projection angles would show no clear evidence of collapse.} In the second column artificial small-scale structure is visible in the sheet and is amplified as it collapses along a second axis, as seen in the third column. It is, however, far weaker than the fragmentation in Fig.~\ref{fig:H1field}. Finally, the halo forms in the rightmost column. The lower panels show the spherically averaged density profile, and by $a=0.45$ (rightmost column), the halo has developed a $\rho\simeq 90 r^{-3/2}$ inner density cusp.

So far we have only studied haloes arising from the $n=1$ power spectrum. We now show examples of haloes arising from each of the $n=-2$ and $n=-2.67$ power spectra. Figures \ref{fig:W1field} and~\ref{fig:C1field} show the density fields surrounding haloes W1 and C1, respectively, during their collapse. Each exhibits its own unique behaviour. The material of the W1 halo (Fig.~\ref{fig:W1field}) initially collapses into a filament, but the halo forms at the end of this filament, so unlike H1 (Fig.~\ref{fig:H1field}), it is not clear that W1 can be viewed as arising from the collapse of the filament itself. Meanwhile, the material of the C1 halo (Fig.~\ref{fig:C1field}) exhibits a range of behaviour during its collapse. It first collapses into a very pronounced sheet; this sheet subsequently collapses into a filament followed almost immediately by formation of the halo, which promptly accretes another collapsed filament edge-on. Despite these variations, both haloes develop an inner density profile close to $\rho\simeq 80 r^{-3/2}$ by the final snapshot shown.

Figure~\ref{fig:profiles-coll} shows density profiles during the collapse phase and shortly afterwards for all twelve of our high-resolution haloes. All of these haloes develop density profiles with $\rho\sim r^{-3/2}$ that remain effectively stable over a factor of at least 1.3 in scale factor, extending to larger radii as the haloes accrete new material. Some of these density profiles, such as those of C1 and C2, do grow moderately in amplitude over time, an effect we explore further in Section~\ref{sec:evo}. The density profiles of our highest-resolution haloes -- W5 and C4 -- show some tendency toward shallowing at the smallest radii. However, it is unclear whether this behaviour is converged with respect to simulation resolution, as we discuss in Appendix~\ref{sec:convergence}, and studies with yet higher resolution are needed to test this. Aside from the uncertainty in these two cases, the $\rho\propto r^{-3/2}$ density cusp appears to be a robust consequence of the direct collapse of a smooth initial density peak, independent of the details of the collapse.

\subsection{Connection to the precursor density field}\label{sec:peaks}

Despite the varied details of the collapse process, the central cusps of the first haloes develop from particularly simple initial conditions: the inner regions of smooth peaks in the primordial density field. \citet{2019PhRvD.100b3523D} noted that since the $\rho\propto r^{-3/2}$ cusp stabilizes so quickly after collapse, its properties must be sensitive only to the immediate neighborhood of the precursor density peak. Such a peak can be locally characterized by its height $\delta$ and its characteristic comoving radius $R\equiv |\delta/\nabla^2\delta|^{1/2}$. Its collapse dynamics are also sensitive to the tidal tensor $\partial_i\partial_j\phi$ (where $\phi$ is the peculiar potential), which can be described by its ellipticity $e$ and prolateness $p$.
\chg{Specifically, if $\lambda_1>\lambda_2>\lambda_3$ are the eigenvalues of $-\partial_i\partial_j\phi$, then $e\equiv \frac{\lambda_1-\lambda_3}{2(\lambda_1+\lambda_2+\lambda_3)}$ and $p\equiv \frac{\lambda_1+\lambda_3-2\lambda_2}{2(\lambda_1+\lambda_2+\lambda_3)}$. Note that since $\lambda_1+\lambda_2+\lambda_3\propto \delta$, the three quantities $\delta$, $e$, and $p$ fully specify the tidal tensor up to rotation.}

From dimensional considerations, we expect that the amplitude $A$ of the $\rho = A r^{-3/2}$ cusp should be proportional to $\bar\rho(a_c) (a_c R)^{3/2}$, where $a_c$ is the expansion factor at which the peak collapses and $\bar\rho(a_c)=\bar\rho a_c^{-3}$ is the cosmological density at $a_c$. As above, $\bar\rho$ is the comoving cosmological density (equal to 1 in our simulations). Consequently, we may write
\begin{equation}\label{A}
    A = \alpha \bar\rho a_c^{-3/2} R^{3/2}
\end{equation}
for some universal proportionality constant $\alpha$. Here we evaluate $a_c$ as a function of $\delta$, $e$, and $p$ using the ellipsoidal collapse approximation in \citet{2001MNRAS.323....1S}.
\chg{This approximate $a_c$ lies within about 5 per cent of the scale factor at which the $\rho\propto r^{-3/2}$ cusp first appears in each simulation, and using the latter scale factor to predict $A$ does not appreciably alter the degree to which it matches simulation results.}

We can also estimate the size of the $\rho\propto r^{-3/2}$ central cusp shortly after collapse. Close to a peak in the linear density field, the mean density contrast within comoving radius $q$ scales as $1-(q/R)^2/10$.\footnote{This is the mean overdensity {\it within} $q$; the mean overdensity {\it at} $q$ scales as $1-(q/R)^2/6$.} Spherical shells at radii $q>0$ can be approximated as undergoing ellipsoidal collapse within the same tidal field with this lower enclosed density contrast. Now consider the shell that collapses at $a=1.17a_c$. A factor of 1.17 in $a$ corresponds to one dynamical time, in the sense of \citet{2008gady.book.....B}, for an object whose density is 200 times the cosmological mean. This shell has Lagrangian radius $q$ satisfying $1-(q/R)^2/10 = 1.17^{-1}$, so it encloses the mass
\begin{equation}\label{M0}
    M_0
    = \frac{4\pi}{3} q^3\bar\rho
    = \frac{4\pi}{3}10^{3/2}\left(1-\frac{1}{1.17}\right)^{3/2}R^3 \bar\rho
    \simeq 7.3 R^3 \bar\rho.
\end{equation}
The collapsed halo therefore achieves a mass of approximately $M_0$ after one dynamical time interval. If its density profile is $\rho=A r^{-3/2}$ out to the physical radius $r_0$ that encloses $M_0$, then
\begin{equation}\label{r0}
    r_0
    = \frac{1}{4}\left(\frac{3}{\pi}\right)^{2/3} \left(\frac{M_0}{A}\right)^{2/3}
    \simeq 0.92 \alpha^{-2/3} a_c R.
\end{equation}
If the $\rho\propto r^{-3/2}$ cusp is established over roughly the first dynamical time interval after collapse, then $r_0$ may be interpreted as an estimate of the cusp's initial radius.

Within each of our haloes, we identify all of the particles that lie at a radius $r<4.2\times 10^{-3}$ at some time shortly after halo formation (the specific time makes no difference), and we locate the Lagrangian centre of mass of these particles in the initial conditions. We then follow the local density gradient to find the nearest peak in the initial density field. We record the peak height $\delta$ and use Fourier methods to evaluate $R$, $e$, and $p$. We use Eqs. (\ref{A}), (\ref{M0}), and~(\ref{r0}) to evaluate the predicted values of $A$, $M_0$, and $r_0$, respectively, for the associated halo. We assume the proportionality constant $\alpha=24$.\footnote{The value of $\alpha$ that we assume is about 10 per cent smaller than the value that \citet{2019PhRvD.100b3523D} obtained. It is likely that the proportionality constant obtained in that work was biased upward by haloes' later evolution (see Section~\ref{sec:evo}), but since our own halo sample is small and not necessarily representative, the value that we use here might not be more accurate.} Table~\ref{tab:peaks} lists these data. In Fig.~\ref{fig:profiles-coll}, we mark the predicted $A$ for each halo with a horizontal black line, and we terminate that line with a point at the radius $r_0$. We find that with $\alpha=24$, the predicted values of $A$ match the haloes' density profiles mostly very well, albeit with minor scatter. The radius $r_0$ also reasonably estimates the initial size of the $\rho\propto r^{-3/2}$ cusp, although this size cannot be precisely defined.

\begin{table}
	\centering
	\caption{Properties of our haloes' precursor peaks in the linear density field $\delta(\vx)$. The ``offset'' is the comoving distance between the central cusp's mean Lagrangian position and the density peak. $R\equiv |\delta/\nabla^2\delta|$ is the characteristic size of the peak and $\delta(a)/a$ is its height (scaled by $a^{-1}$ since $\delta(a)\propto a$). $e$ and $p$ are the ``ellipticity'' and ``prolateness'' of the tidal tensor $\partial_i\partial_j\phi$ at the location of the peak (see \chg{the text}). Finally, $a_c$ is the peak's predicted collapse time \citep[using the approximation in][]{2001MNRAS.323....1S}, $A$ is the resulting halo's predicted cusp amplitude (Eq.~\ref{A} with $\alpha=24$), and $M_0$ and $r_0$ are the predicted mass and radius of the cusp after one dynamical time interval, respectively (Eqs. \ref{M0} and~\ref{r0}).}
	\label{tab:peaks}
	\tabcolsep=0.16cm
	\begin{tabular}{lccccccccc}
\hline
halo & offset & $R$ & $\frac{\delta(a)}{a}$ & $e$ & $p$ & $a_c$ & $A$ & $M_0$ & $r_0$ \\
\hline
H1 & $0.31$ & $0.96$ & $3.40$ & $0.09$ & $-0.06$ & $0.54$ & $57$ & $6.6$ & $0.058$\\
H2 & $0.54$ & $1.20$ & $3.32$ & $0.18$ & $0.00$ & $0.60$ & $68$ & $13$ & $0.079$\\
H3 & $0.31$ & $1.08$ & $4.09$ & $0.06$ & $0.00$ & $0.43$ & $96$ & $9.3$ & $0.051$\\
W1 & $0.25$ & $0.82$ & $4.86$ & $0.04$ & $-0.02$ & $0.36$ & $84$ & $4.1$ & $0.032$\\
W2 & $0.19$ & $0.89$ & $5.00$ & $0.10$ & $-0.01$ & $0.37$ & $91$ & $5.2$ & $0.036$\\
W3 & $0.00$ & $0.98$ & $4.20$ & $0.09$ & $0.01$ & $0.43$ & $83$ & $6.8$ & $0.046$\\
W4 & $0.11$ & $0.95$ & $4.77$ & $0.06$ & $-0.01$ & $0.37$ & $99$ & $6.3$ & $0.039$\\
W5 & $0.11$ & $0.75$ & $3.48$ & $0.12$ & $-0.01$ & $0.53$ & $40$ & $3.1$ & $0.044$\\
C1 & $0.24$ & $0.96$ & $4.30$ & $0.12$ & $-0.00$ & $0.43$ & $79$ & $6.5$ & $0.046$\\
C2 & $0.11$ & $0.74$ & $4.22$ & $0.07$ & $0.01$ & $0.42$ & $57$ & $3.0$ & $0.034$\\
C3 & $0.00$ & $0.70$ & $4.11$ & $0.10$ & $-0.02$ & $0.44$ & $47$ & $2.5$ & $0.034$\\
C4 & $0.00$ & $0.55$ & $3.66$ & $0.15$ & $-0.07$ & $0.54$ & $25$ & $1.2$ & $0.032$\\
\hline
	\end{tabular}
\end{table}

We also list in Table~\ref{tab:peaks} the offset between the halo cusp's Lagrangian centre and the density peak. In the ideal case where only $\delta$, $R$, $e$, and $p$ determine collapse, this offset should be zero because the halo should form precisely at the density peak. The presence of an offset thus implies that the broader density field around the peak does have a small influence on the collapse process (beyond determining the tidal tensor). The influence of the broader density field could also explain some of the scatter in the cusp amplitudes $\rho r^{3/2}$ in Fig.~\ref{fig:profiles-coll} with respect to the predicted values $A$ from Eq.~(\ref{A}). However, we note that since this offset is always significantly smaller than $R$, there is no ambiguity as to whether a halo actually arose from the peak to which we associate it. We also tried evaluating the predicted value of $A$ using the values of $\delta$, $R$, $e$, and $p$ at the location of the Lagrangian centre instead of the peak, but the outcome did not change significantly.

\section{Persistence of the inner cusp}\label{sec:evo}

\begin{figure}
	\centering
	\includegraphics[width=\columnwidth]{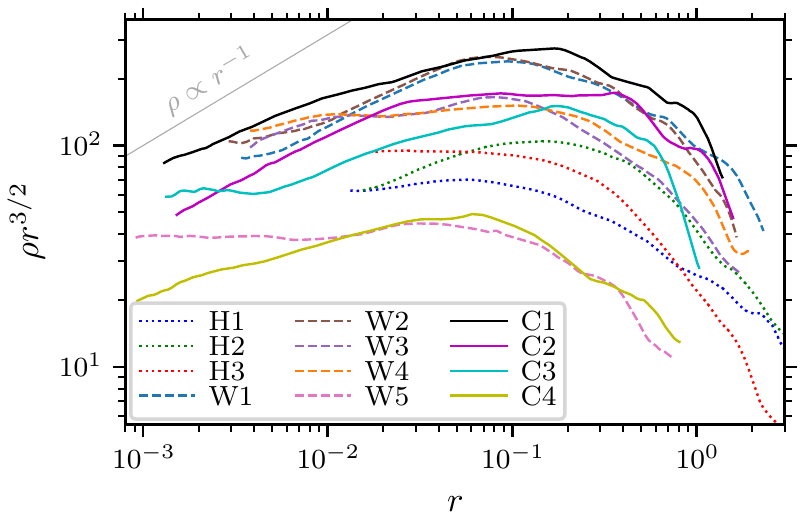}
	\caption{Density profiles of all twelve high-resolution haloes at the end of their simulations (averaged over the final factor of 1.17 in scale factor). We plot these profiles out to the radius $R_{200}$ enclosing an average density 200 times the cosmological mean. While haloes H1, H3, and W5 still retain their $\rho\propto r^{-3/2}$ inner density profiles, most of the profiles have shallowed by this late time. We include a reference $\rho\propto r^{-1}$ line, corresponding to the innermost slope of the NFW profile. Some of the density profiles are beginning to approach this slope.}
	\label{fig:all_late}
\end{figure}

\begin{figure}
	\centering
	\includegraphics[width=\columnwidth]{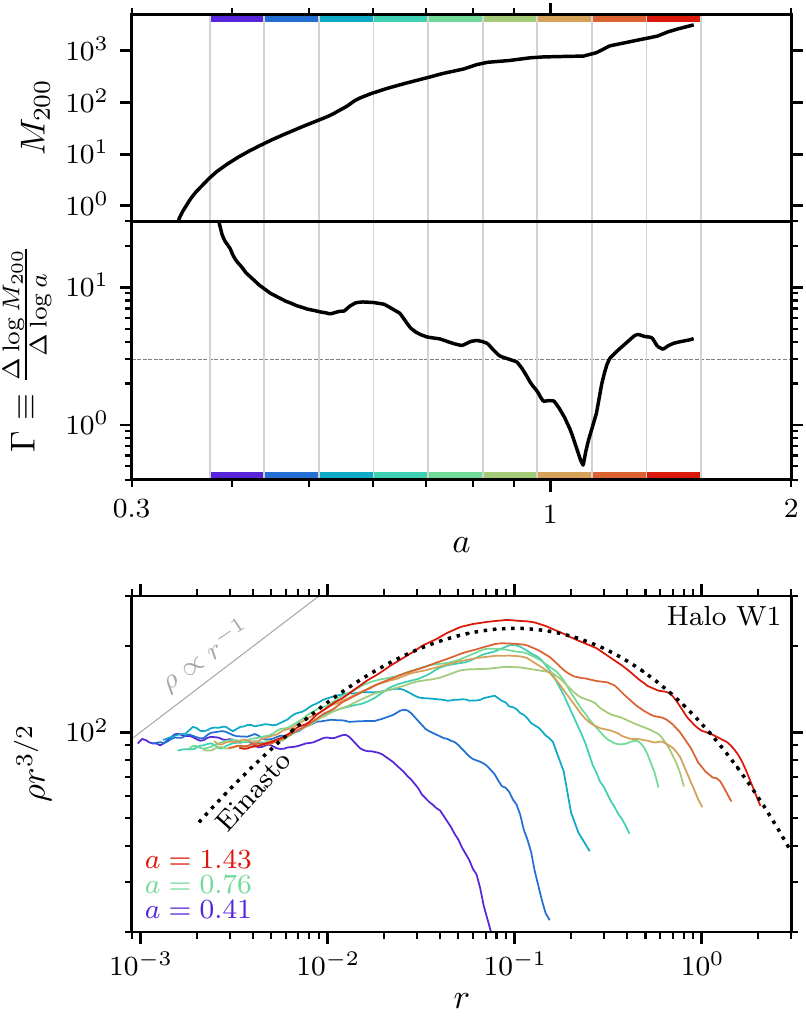}
	\caption{Mass (top), accretion rate $\Gamma$ evaluated at each time over the preceding factor of 1.17 in $a$ (centre), and density profile (bottom) for the W1 halo. We plot the density profile in physical coordinates at a range of different times, indicated by colours; each profile is averaged over the time interval in the upper panels that is marked with the same colour (we also separate these intervals with vertical lines). The halo begins with a $\rho\propto r^{-3/2}$ inner density profile (purple), but its consistently high accretion rate (blue and green) causes the density profile to trend shallower as the halo grows. The accretion rate later drops below $\Gamma\sim 3$ (yellow), and the density profile remains stable for a period. Finally, towards the end of the simulation the accretion rate rises above $\Gamma\sim 3$ again (red), inducing renewed shallowing. At the final time the density profile is well fit by an Einasto profile with $\alpha_\mathrm{E}=0.17$ (Eq.~\ref{einasto}; dotted curve) except at the smallest radii where the natal $\rho\propto r^{-3/2}$ cusp appears to persist. New accretion has mostly built on top of the cusp rather than disrupting it (but see Section~\ref{sec:merge}). A horizontal line marks $\Gamma=3$; the idea that $\Gamma>3$ corresponds to newly accreted material forming a density profile shallower than $\rho\propto r^{-3/2}$ is approximately borne out. In the lower panel we include a $\rho\propto r^{-1}$ line as reference (upper left); $\rho\propto r^{-3/2}$ would correspond to a horizontal line.}
	\label{fig:W1evo}
\end{figure}

\begin{figure*}
	\centering
	\includegraphics[width=\linewidth]{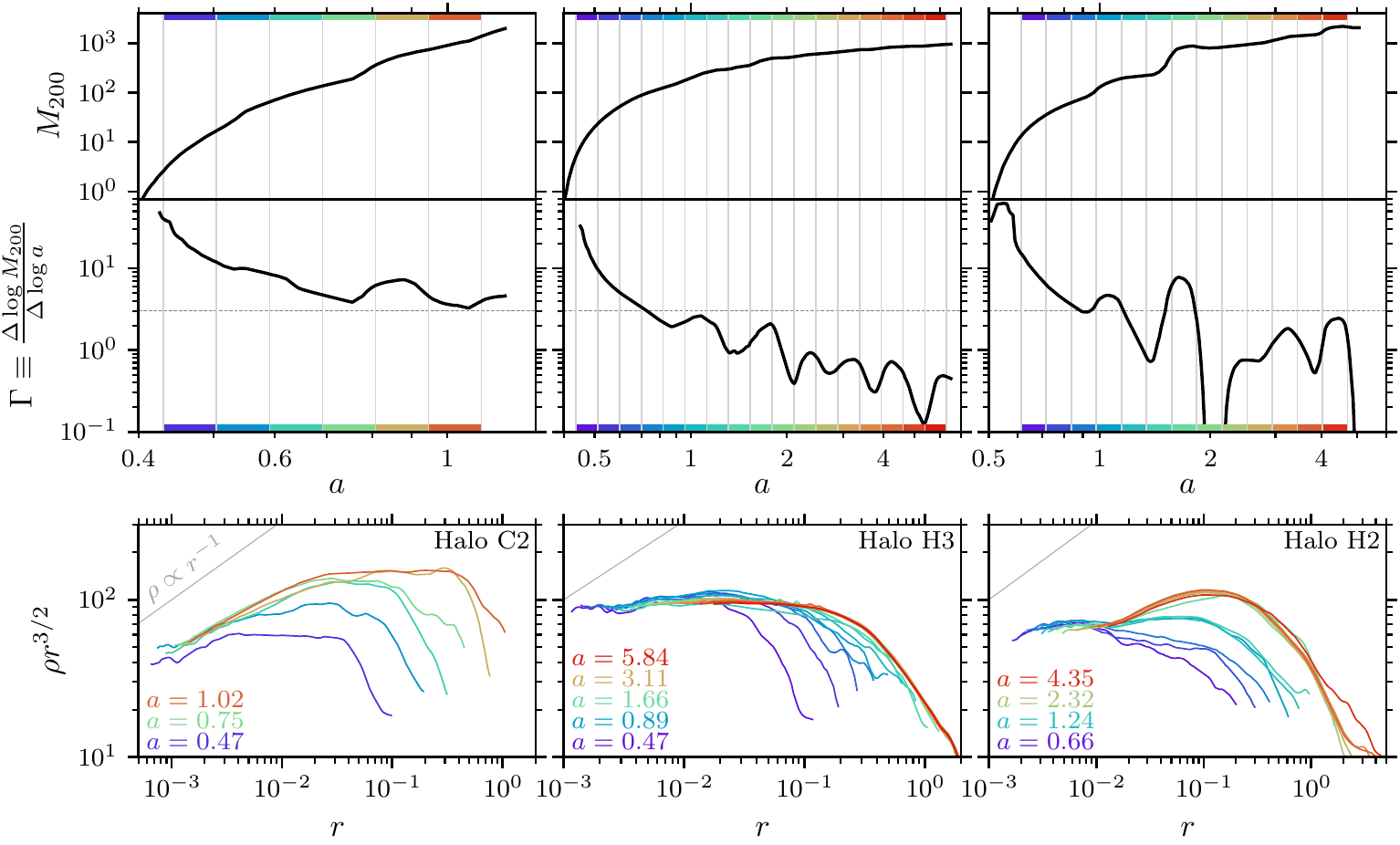}
	\caption{Masses (top), accretion rates $\Gamma$ (centre) and density profiles (bottom) of the haloes C2 (left), H3 (centre), and H2 (right). As in Fig.~\ref{fig:W1evo} we use colours to match the density profiles in each lower panel to the corresponding scale factor band in the accretion-rate plot above it. (See that figure's caption for further explanations.)
	\textit{Left-hand panels}: Halo C2 maintains a high accretion rate $\Gamma$ for some time after formation (like W1 in Fig.~\ref{fig:W1evo}) and its density profile shallows significantly over time, almost reaching $\rho\propto r^{-1}$ by the end of the simulation.
	\textit{Centre panels}: In contrast, halo H3's accretion quickly slows below $\Gamma\sim 3$ and remains low. Its inner $\rho\propto r^{-3/2}$ density profile consequently remains stable over nearly an order of magnitude in scale factor. The accretion rate is so low at late times that even the outer, steeper part of the density profile remains stable from $a\simeq 2$ to $a\simeq 5$.
	\textit{Right-hand panels}: Halo H2's accretion rate initially drops below $\Gamma\sim 3$, causing the inner $\rho\propto r^{-3/2}$ profile to persist intact until $a\simeq 1.6$ (green), at which point a short burst of accretion builds up the inner density profile to a somewhat shallower slope. In fact this accretion episode was associated with a triple merger, which also marginally disrupted the central cusp (see Section~\ref{sec:merge}). The accretion rate remains low thereafter, and the density profile correspondingly remains almost static.}
	\label{fig:evo3}
\end{figure*}

We saw in Section~\ref{sec:form} that a halo's $\rho\propto r^{-3/2}$ cusp forms immediately after collapse. In this section, we explore the extent to which these cusps persist over cosmic time. Figure~\ref{fig:all_late} shows the density profile of every halo at the end of its simulation (see Fig.~\ref{fig:halomass}). Specifically, we average the density profile over the final factor of 1.17 in the scale factor (corresponding again to one dynamical time for something 200 times denser than the cosmological mean). Some of these haloes -- H1, H3, and W5 -- evidently maintain their $\rho\propto r^{-3/2}$ density cusps. However, most of the late-time density profiles exhibit some degree of shallowing, with some, like C2, even approaching the $\rho\propto r^{-1}$ asymptote of the NFW form.

\subsection{The impact of accretion rate}\label{sec:accretion}

The shallowing of the inner profile can largely be understood as a consequence of high accretion rates. Due to its high energy and (typically) high angular momentum, newly accreted material tends to contribute significant density to a halo only at large and intermediate radii, close to the orbital apocentres of the newly accreted particles.\footnote{New material tends to contribute almost uniform density at radii much smaller than the apocentres of its particle orbits \citep[e.g.][]{2010arXiv1010.2539D}. However, this contribution is dwarfed by the much higher density already present at those radii. Only close to the apocentres can the new material significantly boost the halo's density.} Consequently, rapid accretion can shallow a halo's profile by boosting the density at intermediate radii.

Figure~\ref{fig:W1evo} shows how this process plays out. We plot halo mass (upper panel) and mass accretion rate $\Gamma\equiv\Delta\log M_{200}/\Delta\log a$ \citep[centre panel; e.g.][]{2014ApJ...789....1D} for W1 as a function of scale factor $a$, where both $\Delta\log M_{200}$ and $\Delta\log a$ are evaluated over the preceding factor of 1.17 in $a$. The same figure also shows (lower panel) the evolution of the halo density profile over the same period; specifically we time average the profile over successive factors of 1.17 in $a$. The colour of each density profile in the lower panel matches the time range of the same colour marked in the upper panels. While the halo initially has $\rho\propto r^{-3/2}$ in the inner regions, it continues to accrete rapidly for some time after its formation (blue and green), which causes the halo's density to grow at radii $r\gtrsim 10^{-2}$. This density growth is sufficient to produce a shallow density slope, $\diff\log\rho/\diff\log r>-3/2$, in the inner regions. The accretion rate later drops (yellow), leading to a period in which the density profile remains stable except near $R_{200}$. Finally, near the end of the simulation the accretion rate again rises, causing further deposition of mass at intermediate radii.

The connection between the density profile slope $\diff\log\rho/\diff\log r$ and the accretion rate $\Gamma\equiv \diff\log M/\diff\log a$ can be understood on an approximate level by means of a simple model. Due to its mass growth, the halo's radius $R$ grows at the rate $\diff \log R/\diff\log a = 1+\Gamma/3$ in physical coordinates. Now suppose that material accreted at the scale factor $a^\prime$ contributes significant density within the halo only near a particular radius $r=x R(a^\prime)$ that is proportional to the halo's radius $R(a^\prime)$ at the accretion time (so $x$ is some fixed number). Then the profile $M(r)$ of the halo's enclosed mass obeys
\begin{equation}\label{selfsimM}
    \frac{\diff \log M(r)}{\diff \log r} 
    = \frac{\diff \log M(a)}{\diff \log a}\left(\frac{\diff \log R(a)}{\diff \log a}\right)^{-1}
    = \frac{3\Gamma}{3+\Gamma}
\end{equation}
at $r=x R(a)$ if $\Gamma$ is evaluated at $a$. Despite this model's simplicity, close variants thereof have been shown to successfully predict trends in the density profiles within halo populations \citep[][]{2010arXiv1010.2539D,2013MNRAS.432.1103L,2019PhRvD.100b3523D}.\footnote{\citet{2010arXiv1010.2539D} and \citet{2019PhRvD.100b3523D} considered models framed in terms of a halo's initial density peak instead of its accretion rate, but the idea is otherwise almost identical. These works also considered variants in which the radius at which accreted material settles is allowed to change in response to later accretion. The model in \citet{2013MNRAS.432.1103L} is equivalent to the claim that Eq.~(\ref{selfsimM}) holds if $\Gamma$ is evaluated at the scale factor $a$ such that $M(a)=M(r)$, where $M(a)$ is the halo's mass at $a$ while $M(r)$ is the enclosed mass profile at the final time. Note that \chg{these models cannot explain the $\rho\propto r^{-3/2}$ cusp.}} Equation~(\ref{selfsimM}) suggests that if a halo accretes material at a rate $\Gamma>3$, then the mass profile resulting from this accretion should obey $\diff\log M/\diff\log r>3/2$ at the corresponding radius, which requires a density profile shallower than $\rho\propto r^{-3/2}$. Thus, we expect that accretion at a rate $\Gamma\gtrsim 3$ should build up a density profile with a logarithmic slope shallower than $-3/2$. This expectation is approximately borne out in Fig.~\ref{fig:W1evo}, where in the centre panel we mark $\Gamma=3$ with a grey line. When $a\lesssim 0.9$ and when $a\gtrsim 1.2$, $\Gamma>3$, which causes the density profile to gradually grow at intermediate radii $10^{-2}\lesssim r\lesssim 10^{-1}$ into a form with a slope shallower than $\rho\propto r^{-3/2}$. In contrast, when $0.9\lesssim a\lesssim 1.2$, $\Gamma<3$, and during this interval the inner and intermediate parts of the density profile remain stable.

Since the primary impact of rapid accretion is to boost the halo's density, the shallowing of the density profile predominantly occurs at intermediate radii $10^{-2}\lesssim r\lesssim 10^{-1}$. The central $\rho\propto r^{-3/2}$ cusp largely survives the process, albeit with some mass loss that we will discuss further in Section~\ref{sec:merge} (where we attribute it to a major merger event). In fact, W1's density profile in the final snapshot is well fit at almost all radii by the Einasto density profile (dotted curve)
\begin{equation}\label{einasto}
    \rho(r)=\rho_{-2}\exp\left\{-\frac{2}{\alpha_\mathrm{E}}\left[\left(\frac{r}{r_{-2}}\right)^{\alpha_\mathrm{E}}-1\right]\right\}
\end{equation}
\citep[][]{1965TrAlm...5...87E}, where $\rho_{-2}$ and $r_{-2}$ are scale parameters defined here as the density and radius at which the profile's logarithmic slope crosses $\diff\log\rho/\diff\log r=-2$. Here we fix $\alpha_\mathrm{E}=0.17$ and find $r_{-2}=0.55$ and $\rho_{-2}=380$.  Equation~(\ref{einasto}) is the same density profile seen in cold dark matter simulations at much larger scales where the free-streaming cutoff is not resolved \citep[e.g.][]{2004MNRAS.349.1039N,2010MNRAS.402...21N}, and such simulations also find that $\alpha_\mathrm{E}\simeq 0.17$ supplies a good fit. Only at the smallest resolved scales, $r\lesssim 5\times 10^{-3}$, does the density profile here break away from the Einasto form due to the persistence of the initial $\rho\propto r^{-3/2}$ central cusp.

Figure~\ref{fig:evo3} shows three more examples of how the accretion rate $\Gamma$ relates to the shallowing of the halo's density profile.
\begin{enumerate}
\item Halo C2 (left) forms with a density profile that is initially close to $\rho\propto r^{-3/2}$, but its accretion rate $\Gamma$ remains high, so this profile rapidly shallows. By the end of the simulation the inner density profile is close to $\rho\propto r^{-1}$.
\item In contrast, the accretion rate of halo H3 (centre) quickly drops after its formation, allowing its inner $\rho\propto r^{-3/2}$ density profile to remain stable over nearly an order of magnitude in scale factor. The accretion rate in this case becomes so low that even the outer, steeper part of the density profile stabilizes after $a\simeq 2$.
\item Halo H2's accretion rate (right) also slows quickly early on, leading to an initially static inner density profile. However, around $a\simeq 1.6$ a brief period of rapid accretion builds up the density profile to a shallower slope. Subsequently the halo's accretion rate remains low, and its density profile correspondingly remains static again.
\end{enumerate}

\subsection{Impact of halo mergers}\label{sec:merge}

\begin{figure}
	\centering
	\includegraphics[width=\columnwidth]{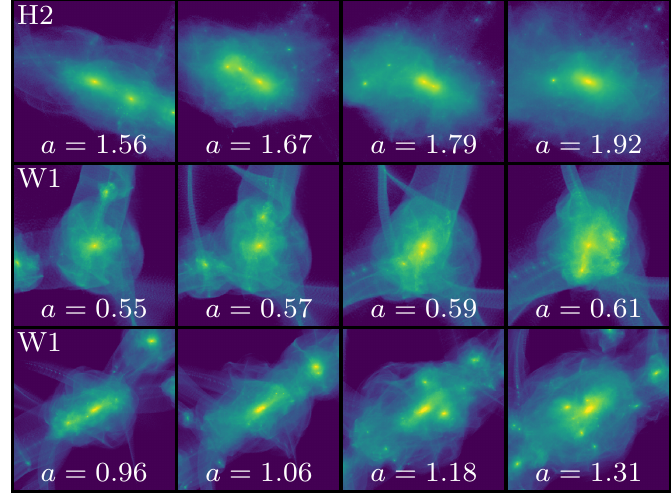}
	\caption{The significant merger events (each a different row) that we consider in Section~\ref{sec:merge}.  From left to right we advance in time.
	\textit{Top row}: H2's merger event at $a\sim 1.6$ is a three-body event where the two smaller haloes have masses of about 1/3 and 1/4 the main progenitor's mass. This event disturbs the halo's established $\rho\propto r^{-3/2}$ density profile by about 10 per cent (see Fig.~\ref{fig:evo3}).
	\textit{Centre row}: W1's merger event at $a\sim 0.6$ is also a three-body event; in this case the two smaller haloes have somewhat smaller masses of about 1/4 and 1/5 the main progenitor's mass. This event disturbs the halo's established $\rho\propto r^{-3/2}$ density profile to a similar degree (see Fig.~\ref{fig:W1evo}).
	\textit{Bottom row}: W1's second major merger event at $a\sim 1.3$ has a mass ratio of about 1/3 but does not appreciably disturb the central $\rho\propto r^{-3/2}$ cusp, at least within our resolution limits.}
	\label{fig:H2merge}
\end{figure}

\begin{figure*}
	\centering
	\includegraphics[width=\linewidth]{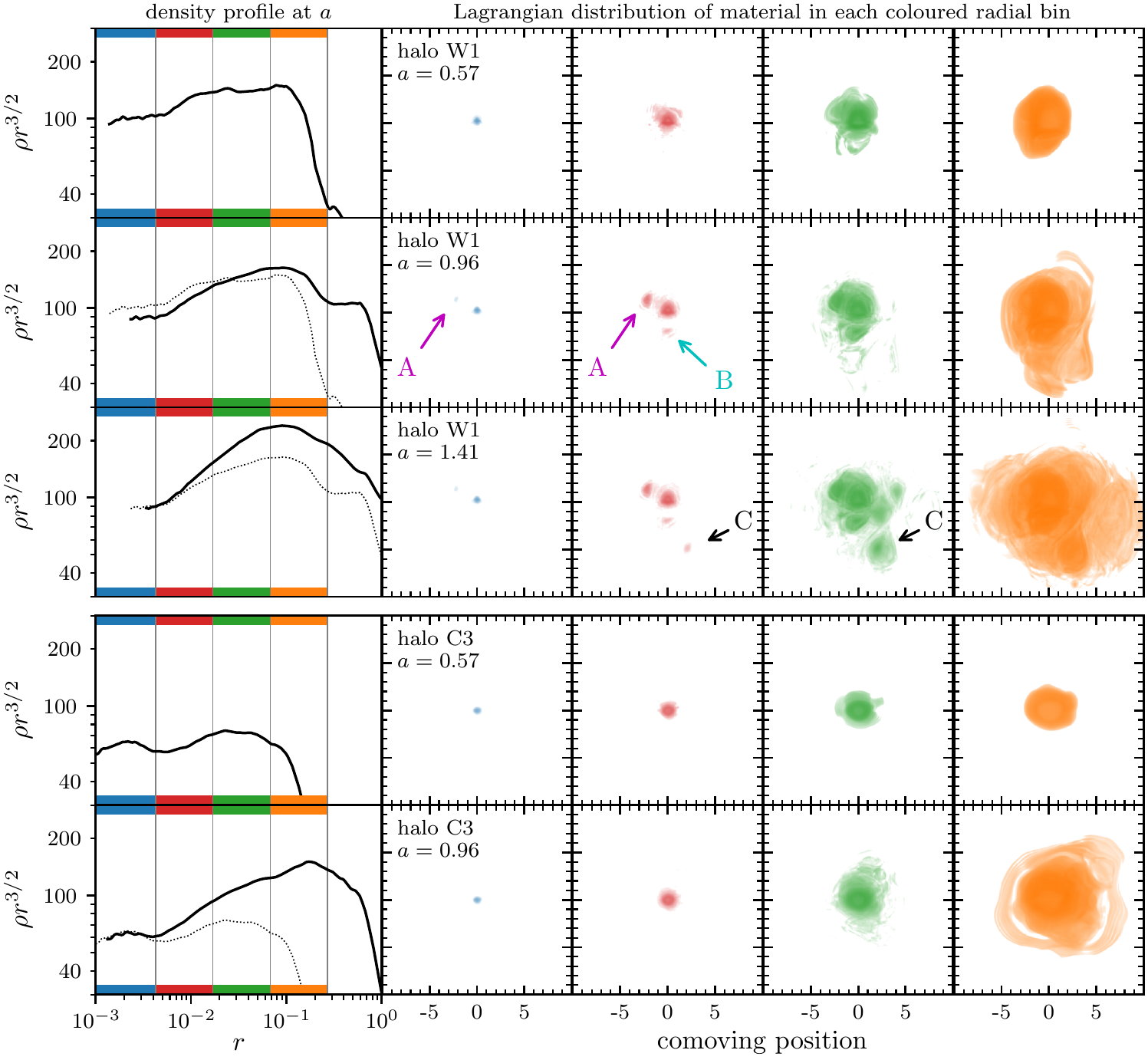}
	\caption{Lagrangian regions associated with halo material at set radii. Within each row, the left-hand panel shows a halo's density profile at a particular time (averaged over a factor of 1.17 in $a$), and in some rows we also overplot the previous row's density profile as a dotted curve for comparison. We mark four coloured radial bins (also separated by vertical lines). The panels to the right of each density profile show the projected distribution of Lagrangian positions (i.e. positions within the initial conditions) associated with the material that resides in the radial bin of the same colour (and we consider the same time interval, distributing each particle across multiple bins if necessary). The colour intensity scale is logarithmic between $10^{-3}$ and $1/4$ the total projected density (over a cubic region). The upper three panels show the W1 halo at different times. At first, the Lagrangian region associated with each radial bin is simply connected, and bins at larger radii  are associated with larger Lagrangian regions. This behaviour indicates that material that accreted later (because it was initially farther from the peak) contributes at larger radii within the halo (note that the red, green and yellow regions in all panels are hollow, although this is difficult to see in projection). The second and third rows of panels for W1 show disconnected Lagrangian regions `A', `B', and `C' contributing to the inner radial bins. Each additional region is associated with a separate halo that merged onto the main system; dynamical friction allowed this these haloes to deposit material close to the centre of W1. The merger events associated with the `A' and `B' regions partially disrupted the halo's central density profile, but the merger associated with the `C' region had no such effect, depositing all its material at intermediate radii. In contrast, the lower two panels show the C3 halo. This halo did not undergo any major mergers, and the Lagrangian distribution of the material in its inner radial bins is simply connected and nearly spherically symmetric at both times. Nevertheless, rapid accretion of diffuse material caused this halo to develop a shallow density profile at all but the innermost radii, where the initial $\rho\propto r^{-3/2}$ cusp is preserved.}
	\label{fig:origin}
\end{figure*}

\begin{figure*}
	\centering
	\includegraphics[width=\linewidth]{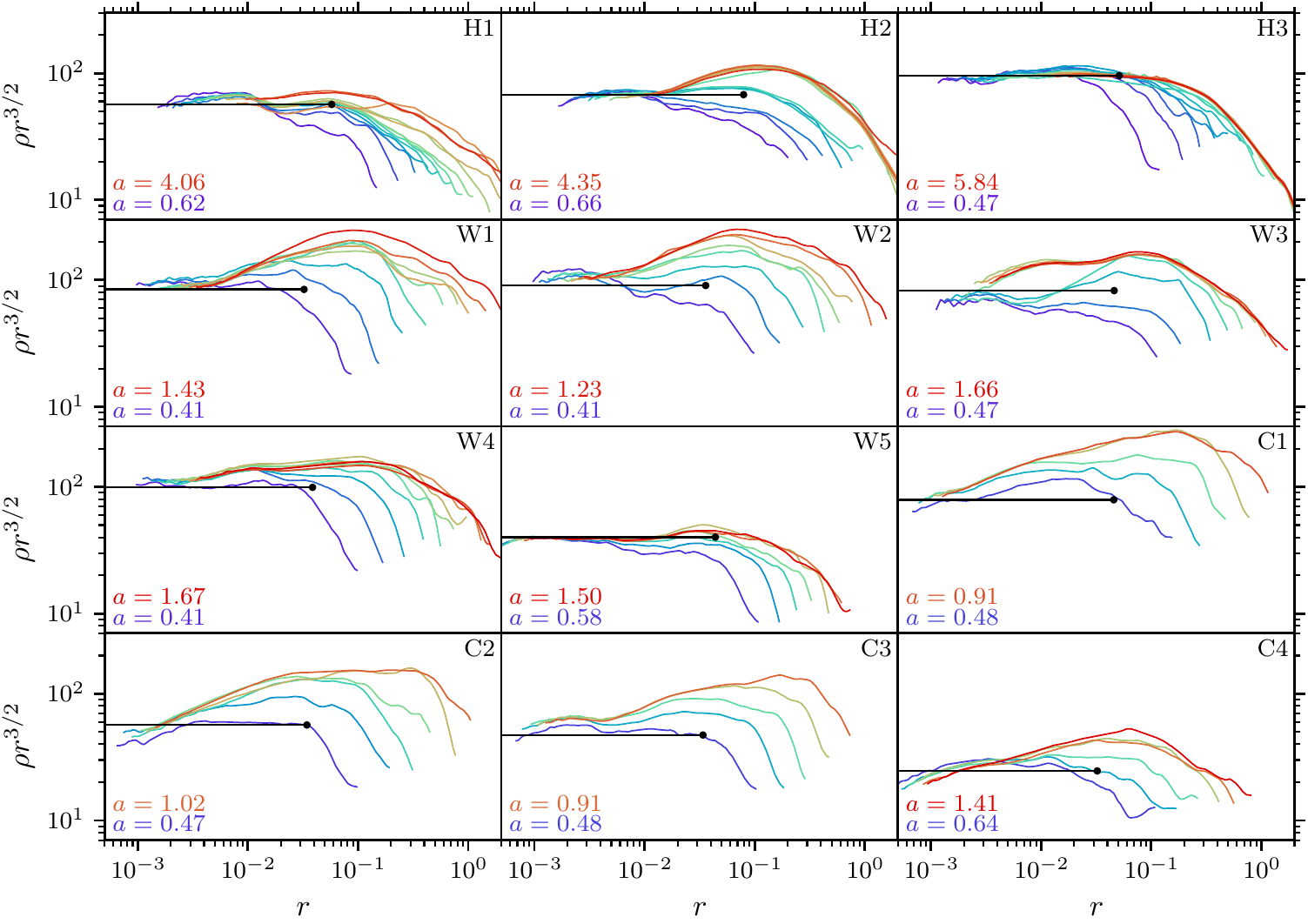}
	\caption{Density profile evolution for all twelve haloes. As in earlier figures, the density profiles are time averaged over successive factors of $1.17$ in $a$ (different colours). We find no evidence, within our resolution limits, for significant disruption of the natal cusps due to merger events. The effects are limited to a 10 to 20 per cent reduction in amplitude. We also repeat the cusp predictions (black lines) from Fig.~\ref{fig:profiles-coll}, which are based on each halo's precursor density peak. Even after halo growth, the initial peak continues to predict the inner cusp reasonably well in cases where the cusp persists within our resolution limits.}.
	\label{fig:evo12}
\end{figure*}

\citet{2016MNRAS.461.3385O} and \citet{2017MNRAS.471.4687A} showed through controlled merger simulations that major mergers (between haloes of comparable masses) can disrupt $\rho\propto r^{-3/2}$ central cusps. We now present a couple of examples that confirm this effect in a cosmological setting.
\begin{itemize}
    \item Halo H2 merged at $a\simeq 1.6$ with two other haloes of roughly 1/3 and 1/4 the main progenitor's mass. This event is pictured in the upper row of panels in Fig.~\ref{fig:H2merge}. The right-hand panels of Fig.~\ref{fig:evo3} show that the amplitude of the central $\rho\propto r^{-3/2}$ cusp decreased by about 10 per cent at this time (and its slope may have also changed). The large influx of mass mostly settled at larger radii, boosting the density profile around $r\sim 10^{-1}$.
    \item Halo W1 merged at $a\simeq 0.6$ with two other haloes of masses roughly 1/4 and 1/5 the main progenitor's mass; this event is pictured in the central row of panels in Fig.~\ref{fig:H2merge}. Figure~\ref{fig:W1evo} shows that like H2's merger, this event reduced the amplitude of the central density cusp around $r\sim 10^{-2}$ by about 10 per cent (although we do not have the resolution to determine whether the slope also changed). Here, too, the mass primarily settled at larger radii, building up the density profile around $r\sim 10^{-1}$.
\end{itemize}

Major mergers have the capacity to alter a halo's structure at small radii because dynamical friction causes the core of a large subhalo to sink rapidly to the centre of a host, shedding most of its mass along the way \citep[e.g.][]{2010gfe..book.....M}. The subhalo heats the host's material in the process. This behaviour contrasts with that of smoothly accreted material and minor mergers, which do not sink in this way and therefore impact the growing  halo's structure only at intermediate to large radii (see Section~\ref{sec:accretion}). We illustrate this effect in Fig.~\ref{fig:origin}. This figure shows a halo's density profile at a particular time in the left-hand panels, and we mark radial bins within the density profile with different colours. The panels to the right of each density profile show the distribution of Lagrangian positions (within the initial conditions) of material residing within each coloured radial bin. The uppermost row of panels shows halo W1 at an early time, before any significant mergers have taken place. The Lagrangian regions associated with each radial bin of the density profile are simply connected and nested at this time, with shells of larger (final) radius associated with larger Lagrangian regions. This behaviour indicates that material that was initially at a larger radius within the halo's precursor density peak -- and therefore accreted later -- contributes primarily at larger radii within the non-linear halo cusp. In contrast, the second row of panels shows halo W1 after the set of major mergers discussed above has taken place. Disconnected Lagrangian regions, which we label `A' and `B', now contribute even within the smallest radial bins (blue and red). The inner regions of W1 thus contain material from both merging haloes; both contribute to the final profile at all radii, even to the inner cusp. The amplitude of the density profile of W1 is decreased at small radii as a result of this merging (the density profile from the uppermost row is overplotted here as a dotted curve for comparison).

However, disruption of the central cusp is not a guaranteed outcome of a major merger. In fact, W1 underwent another merger event at $a\simeq 1.3$ with a mass ratio of about $1/3$, an event that is pictured in the lower panels of Fig.~\ref{fig:H2merge}. The disconnected Lagrangian region associated with this merging halo is labeled `C' in the third row of panels in Fig.~\ref{fig:origin}. This halo's material was able to contribute density as deep as the red radial bin within W1. However, this event did not appreciably disturb the innermost part of the density profile, at least within our resolution limits. As before, we overplot the previous row's density profile as a dotted curve.

We can also use plots of this kind to show explicitly that major mergers are not necessary to produce a shallow density profile. In the lower two rows of Fig.~\ref{fig:origin}, we show the density profile of halo C3 at two different times. As before, the right-hand panels plot the Lagrangian positions of the material that resides within each coloured radial bin at each time. This halo evidently develops a density profile at intermediate radii that is significantly shallower than $\rho\propto r^{-3/2}$. However, there is no significant disconnected Lagrangian region that would mark the impact of a major merger event.\footnote{The ring-like structure visible in the bottom-right panel is associated with coherence of the orbital phases of freshly accreted material. Material initially within the ring is close to the first pericentre of its orbit through the halo, while material initially in the white region just inside the ring is close to its first apocentre subsequent to infall and currently lies outside the orange radial range in the profile panel. Note that halo C3 formed from a filament, and the Lagrangian ring is aligned along that filament's axis.} Instead, material at large Lagrangian distances (which was accreted later) remains at large radii within the halo. Accordingly, the central $\rho\propto r^{-3/2}$ cusp remains undisturbed. Indeed, halo C3's most major merger event was associated with a mass ratio of about $1/20$.

The degree to which mergers disrupt the natal cusps of our haloes appears at most moderate. Figure~\ref{fig:evo12} shows the density profile evolution for all twelve of our haloes. Despite the significant shallowing of many density profiles at intermediate radii -- and sometimes even at the smallest resolved radii -- due to deposition of accreted material, we find no case where the initial cusp suffered more than 10 to 20 per cent suppression. Resolution limitations mean that there are cases -- particularly H2, W3, C1, C2, and C4 -- where we cannot exclude that more major disruption occurred at the smallest radii. Nevertheless, in no case do we find positive evidence for substantial disruption of the central cusp. In cases where the central $\rho=A r^{-3/2}$ cusp persists within our resolution limits, the properties of the halo's precursor peak continue to predict its coefficient $A$ with reasonable accuracy (horizontal black lines; see Section~\ref{sec:peaks}).

\subsection{Mass and size of the inner cusp}\label{sec:size}

\begin{figure}
	\centering
	\includegraphics[width=\columnwidth]{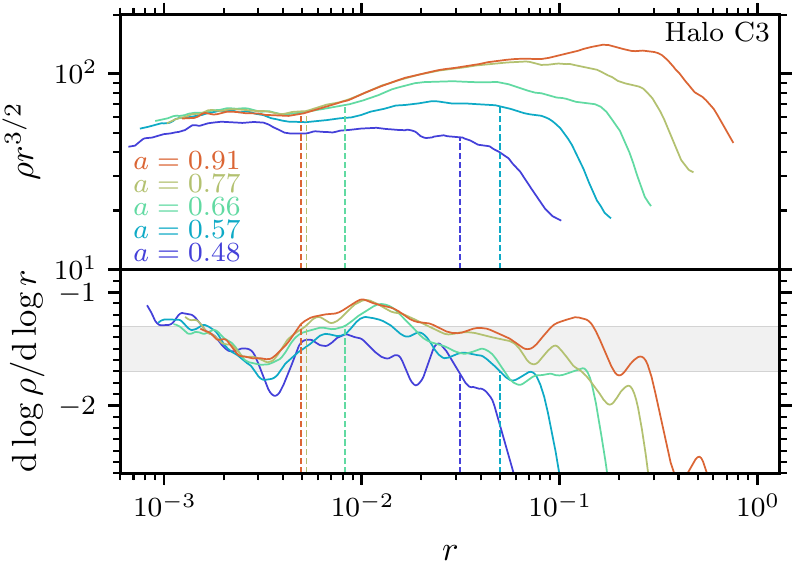}
	\caption{Demonstration of the procedure through which we identify the radius $R_\mathrm{cusp}$ of the central cusp. We plot the density profile (top) and logarithmic density slope $\diff\log\rho/\diff\log r$ (bottom) of the C3 halo at a number of different times (colours). At each time, we identify the cusp radius $R_\mathrm{cusp}$ (vertical line) as the largest radius at which $\diff\log\rho/\diff\log r$ remains within the range (-1.7,-1.3) (grey band), but we ignore small excursions; see the text. The resulting radius values appear to reasonably mark the outer edge of the $\rho\propto r^{-3/2}$ cusp.}
	\label{fig:slope}
\end{figure}

\begin{figure}
	\centering
	\includegraphics[width=\columnwidth]{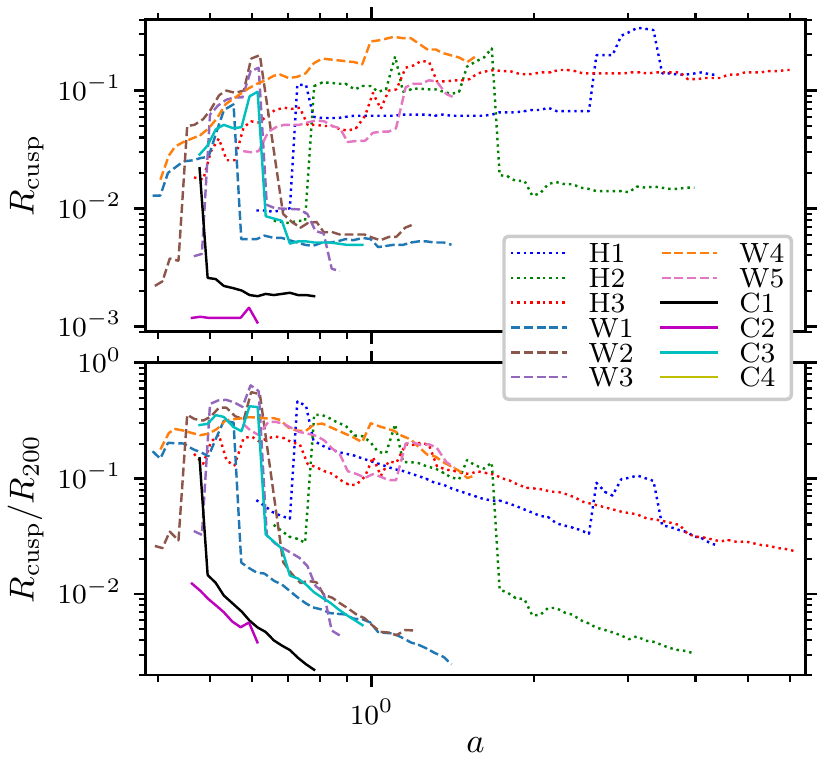}
	\caption{Evolution of the cusp radii $R_\mathrm{cusp}$ of our haloes as a function of the expansion factor $a$. The upper panel shows $R_\mathrm{cusp}$ in absolute units; generally we find that $R_\mathrm{cusp}$ grows initially, drops suddenly at one time as rapid accretion begins to build up a shallower density profile, and otherwise remains stable. The lower panel shows instead $R_\mathrm{cusp}/R_{200}$, i.e. the cusp radius in units of the halo's radius $R_{200}$ (enclosing 200 times the cosmological mean density). Rapidly accreting haloes arising from the $n=-2.67$ and $n=-2$ power spectra quickly grow to the extent that $R_\mathrm{cusp}/R_{200}< 10^{-2}$, but haloes W4, W5, H1, and H3 maintain slower accretion rates such that $R_\mathrm{cusp}$ remains close to $10^{-1}R_{200}$.}
	\label{fig:Rcusp}
\end{figure}

\begin{figure}
	\centering
	\includegraphics[width=\columnwidth]{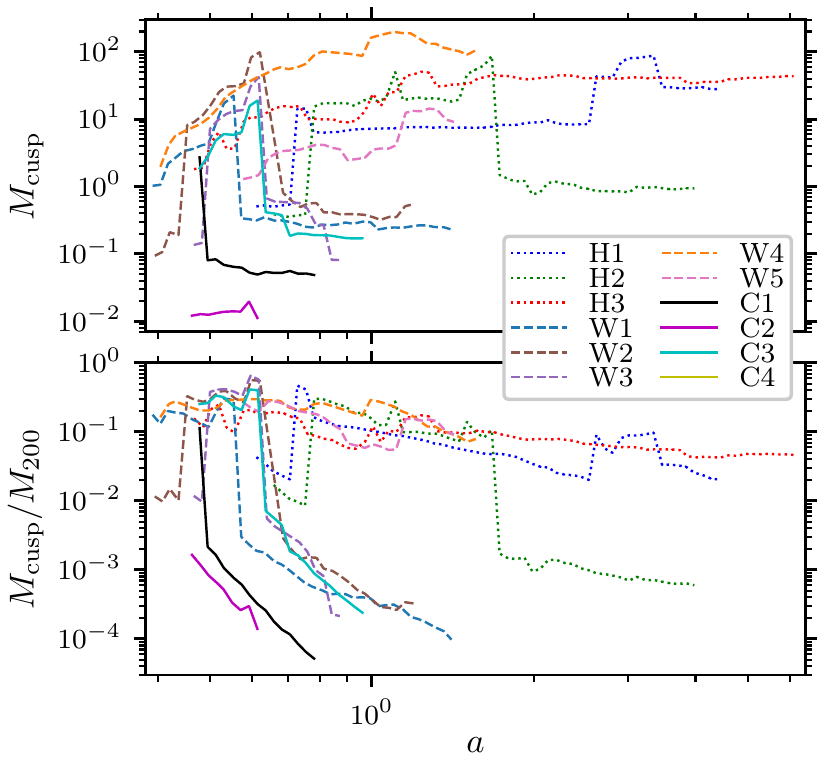}
	\caption{Evolution of the masses $M_\mathrm{cusp}$ enclosed within our haloes' cusp radii $R_\mathrm{cusp}$. The upper panel shows $M_\mathrm{cusp}$ in absolute units; behaviours similar to those in Fig.~\ref{fig:Rcusp} are evident. The lower panel shows $M_\mathrm{cusp}/M_{200}$, i.e. the cusp mass in units of the halo's mass $M_{200}$ (enclosed within $R_{200}$). We find that the rapidly accreting haloes C1--C4 and W1--W3 quickly grow to such extent that $M_\mathrm{cusp}/M_{200} \lesssim 10^{-3}$. On the other hand, slowly accreting haloes like H1, H3, W4, and W5 are able to maintain $M_\mathrm{cusp}/M_{200}\sim 10^{-1}$.}
	\label{fig:Mcusp}
\end{figure}

\begin{figure}
	\centering
	\includegraphics[width=\columnwidth]{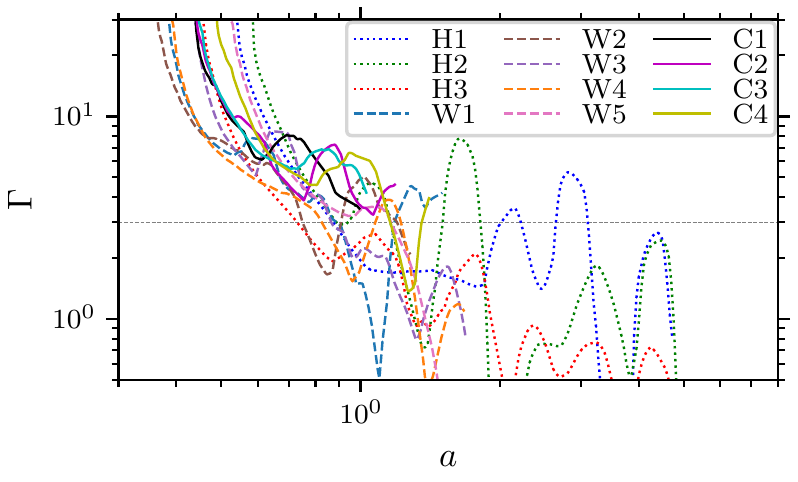}
	\caption{Accretion rates $\Gamma\equiv \Delta\log M_{200}/\Delta\log a$ for our haloes, evaluated as in Section~\ref{sec:accretion}. The horizontal reference line marks $\Gamma=3$.}
	\label{fig:halogamma}
\end{figure}

We have now seen many examples of how a halo's density profile evolves after formation of its initial $\rho\propto r^{-3/2}$ cusp. For instance, halo W1 (Fig.~\ref{fig:W1evo}) rapidly built a shallow Einasto density profile on top of its initial cusp, so that the latter only persists at radii $r\lesssim R_{200}/400$. Halo C2 (Fig.~\ref{fig:evo3}) underwent similar evolution to the extent that its central cusp (if it persists) is not even resolved at the final snapshot. In contrast, halo H3's $\rho\propto r^{-3/2}$ cusp remains a significant portion of the halo even at the end of the simulation. We now quantify more precisely how the sizes of our haloes' natal cusps evolve over time.

We define the size and mass of the $\rho\propto r^{-3/2}$ cusp in the following manner. First, we evaluate $\diff\log\rho/\diff\log r$ at each radius $r$ by carrying out a linear regression in log space over the surrounding factor of 1.5 in $r$ (that is, radii $r^\prime$ satisfying $r/\sqrt{1.5}<r^\prime<\sqrt{1.5}r$). Next, we search for the smallest radius at which $\diff\log\rho/\diff\log r$ deviates from $-1.5$ by more than $0.2$, but since $\diff\log\rho/\diff\log r$ is noisy, we ignore deviations that are confined to less than a factor of 2 in radius. We define the deviation radius obtained in this way to be $R_\mathrm{cusp}$, the radius of the natal cusp. We demonstrate the outcome of this process in Fig.~\ref{fig:slope} for the C3 halo. It does indeed place $R_\mathrm{cusp}$ where one might reasonably decide that the central cusp ends.

For each halo and each scale factor $a$, we average the density profile over the surrounding factor of $1.17$ in $a$ (that is, scale factors $a^\prime$ satisfying $a/\sqrt{1.17}<a^\prime<\sqrt{1.17}a$) and then evaluate $R_\mathrm{cusp}$ by the above procedure.\footnote{We also make a further adjustment to clean up the evolution plots. Namely, we iteratively remove the data at a scale factor $a$ if the enclosed mass $M_\mathrm{cusp}$ at that scale factor is larger by at least a factor of 3 than at both of its neighboring scale factors. We do the same for scale factors at which $M_\mathrm{cusp}$ is smaller by the same factor than both of its neighbors. Any data points removed in this way are instead interpolated between the neighboring points. This process suppresses a few instances in which $R_\mathrm{cusp}$ and $M_\mathrm{cusp}$ jump wildly. Note that we use all snapshot scale factors, which are separated by roughly a factor of 1.035 in $a$, and not only the scale factors shown in Fig.~\ref{fig:slope}, so we do not remove a significant fraction of the data points.} The resulting evolution of $R_\mathrm{cusp}$ is plotted in Fig.~\ref{fig:Rcusp}. Note that $R_\mathrm{cusp}$ is not always resolved if the slope $\diff\log\rho/\diff\log r$ is already much shallower than -1.5 at the minimum resolved radius. If $R_\mathrm{cusp}$ is not resolved at some time, we do not attempt to resolve it at any future time; attempting to do so may inappropriately identify the central cusp with, e.g., the portion of the Einasto density profile for which the density slope is close to $-1.5$. When $R_\mathrm{cusp}$ ceases to be resolved, we instead simply cut off its time evolution curve in Fig.~\ref{fig:Rcusp}.

In the upper panel of Fig.~\ref{fig:Rcusp}, we see that the physical cusp radius $R_\mathrm{cusp}$ often remains close to constant in time. This behaviour holds during periods of slow accretion but also during rapid accretion as long as accretion builds up the density profile primarily at radii outside of the central cusp. For instance, halo W1's cusp radius remains essentially constant over the time interval $0.6\lesssim a\lesssim 1.4$ even though accretion during that interval is sometimes rapid enough to build up a shallow density profile and sometimes not; compare Fig.~\ref{fig:W1evo}. However, $R_\mathrm{cusp}$ also typically undergoes one (and only one) sudden drop when rapid accretion begins to build on top of an established $\rho\propto r^{-3/2}$ cusp. A prime example of this behaviour occurs for halo H2 at $a\simeq 1.8$; compare the right-hand panels of Fig.~\ref{fig:evo3}. Finally, at early times there can be growth in $R_\mathrm{cusp}$ as the initial collapse builds the density profile outward; compare for instance halo W1's cusp growth during $0.4\lesssim a\lesssim 0.55$ to the plot of its density in Fig.~\ref{fig:profiles-coll} over the same time period. In some cases, growth of $R_\mathrm{cusp}$ can proceed for some time. This outcome is likely attributable to an accretion rate that remains close to $\Gamma=3$ for a while; for example, compare H3's cusp growth during $0.7\lesssim a\lesssim 1.2$ to the contemporaneous accretion rate and density profile in the centre panels of Fig.~\ref{fig:evo3}. In this case, the $\rho\propto r^{-3/2}$ cusp is not entirely set by the initial collapse but is further built outward by later accretion.

\begin{figure*}
	\centering
	\includegraphics[width=\linewidth]{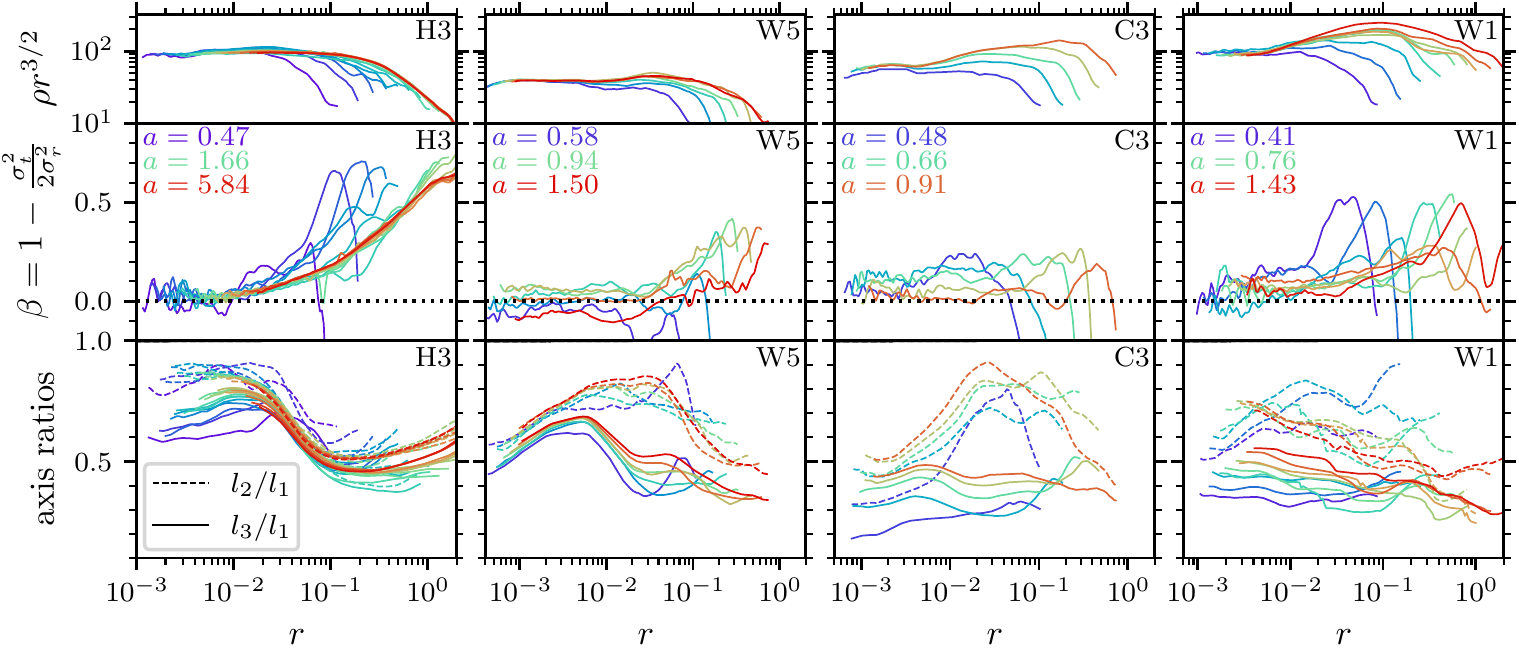}
	\caption{Velocity anisotropy parameter (centre panels) and ellipsoidal axis ratios (lower panels) for the haloes H3, W5, C3, and W1 (from left to right). We also show the density profile at the top, for reference, and all quantities are averaged over successive factors of 1.17 in $a$. The $\rho\propto r^{-3/2}$ central cusps of these haloes have nearly isotropic velocity distributions ($\beta$ close to 0). However, there is far less uniformity in the three-dimensional shapes. H3 and W1 have triaxial cusps that evolve significantly toward a more spherical shape, with H1's cusp becoming nearly spherical at late times. In contrast, W5 and C3 have very prolate central cusps that remain so throughout their evolution, at least at the smallest radii; W5's cusp becomes more oblate at intermediate radii.}
	\label{fig:aniso}
\end{figure*}

The lower panel of Fig.~\ref{fig:Rcusp} plots the evolution of $R_\mathrm{cusp}/R_{200}$ for each halo. For the haloes arising from the $n=-2.67$ power spectrum, accretion is so rapid that the halo's $R_{200}$ radius quickly grows to more than 100 times the cusp radius. This picture demonstrates further why other studies did not find the natal $\rho\propto r^{-3/2}$ cusp inside haloes for which rapid accretion built up a shallower intermediate profile. For instance, \citet{2014ApJ...788...27I} and \citet{2017MNRAS.471.4687A} concluded simply that larger haloes have shallower density profile slopes,\footnote{\citet{2017MNRAS.471.4687A} resolved radii below $10^{-2}R_{178}$ for some of their halo mass bins, and their plot of $\diff\log\rho/\diff\log r$ \citep[Fig. 6 of][]{2017MNRAS.471.4687A} does show signs of steepening at the smallest radii that they plot. Although this effect lies below the radii at which they expect the simulation to be numerically converged, numerical artefacts normally push the density profile towards a shallower rather than a steeper slope (see Appendix~\ref{sec:convergence}).} while \citet{2020Natur.585...39W} found that all haloes are well described by the Einasto fitting form. These studies did not resolve small enough radii to find the natal cusps inside their rapidly accreting haloes.

In Fig.~\ref{fig:Mcusp} we similarly plot the mass $M_\mathrm{cusp}$ enclosed inside $R_\mathrm{cusp}$. For rapidly accreting haloes arising from the $n=-2$ and $n=-2.67$ power spectra, the natal cusp's contribution quickly drops below $10^{-3}$ of the total halo mass. However, haloes W4 and W5 (arising from the $n=-2$ power spectrum) do not accrete rapidly enough to overwhelm the cusp in this way, so they maintain $M_\mathrm{cusp}/M_{200}\sim 10^{-1}$ for the full duration of their respective simulations. The naturally slower accretion rates expected for the $n=1$ power spectrum also cause haloes H1 and H3 to maintain $M_\mathrm{cusp}/M_{200}$ around $10^{-1}$ to $10^{-2}$ during an order-of-magnitude increase in the scale factor $a$. For reference, we show in Fig.~\ref{fig:halogamma} the accretion rates of our haloes; there is a clear correspondence between low accretion rates at late times and large fractional masses in the central cusps.

\subsection{Three-dimensional structure}\label{sec:3d}

\begin{figure*}
	\centering
	\includegraphics[width=\linewidth]{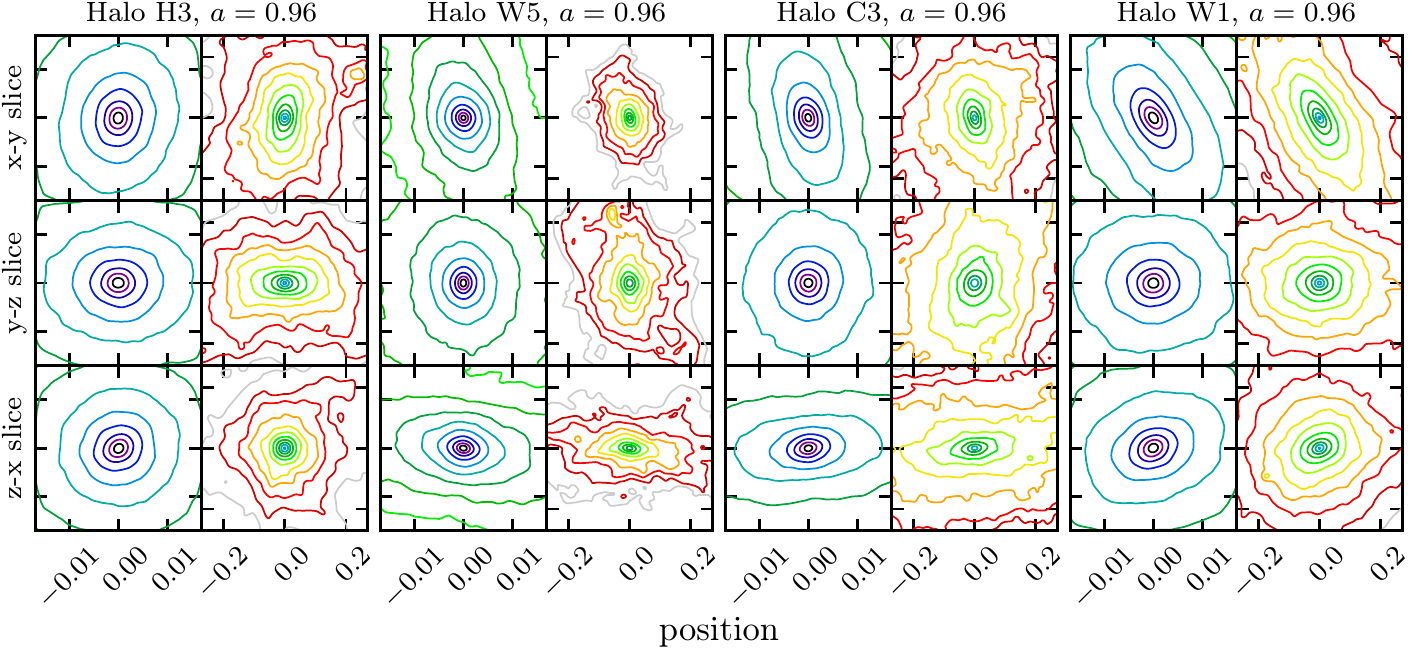}
	\caption{Slices of the equidensity surfaces of haloes H3, W5, C3, and W1 (from left to right) at $a=0.96$. Specifically, we use the density field averaged over the surrounding factor of 1.17 in $a$. The upper, centre, and lower panels show different perpendicular slices, while for each halo, the left-hand and right-hand panels show two different length scales. The equidensity surfaces inside the central cusps (dark curves in each halo's left-hand panels) tend to be aligned with those at larger radii (each halo's right-hand panels).}
	\label{fig:shape}
\end{figure*}

We briefly discuss the three-dimensional structures of our haloes and their evolution over time. We first explore anisotropy in the velocity structure. The centre panels of Fig.~\ref{fig:aniso} show the velocity anisotropy parameter
\begin{equation}
    \beta\equiv 1-\sigma_t^2/(2\sigma_r^2)
\end{equation}
as a function of radius within each of haloes H3, W5, C3, and W1, where $\sigma_r$ and $\sigma_t$ are the radial and tangential velocity dispersions, respectively.
\chg{We evaluate $\sigma_r^2=\langle v_r^2\rangle - \langle v_r\rangle^2$ and $\sigma_t^2=\langle v^2\rangle - \langle v_r^2\rangle$, where $v_r$ is a particle's radial velocity, $v$ is its total velocity, and the angle brackets denote the average over particles in a narrow radial bin centred at radius $r$.}

$\beta=0$ represents an isotropic velocity dispersion, while $\beta=1$ implies that all motion is radial. Generally, \chg{Fig.~\ref{fig:aniso} shows that} $\beta$ grows with radius at small and intermediate radii but drops again as we approach the halo's outer edge. This pattern has been seen in cold dark matter haloes as well \citep[e.g.][]{2010MNRAS.402...21N}. In the context of the first haloes, the interesting feature is that the velocity structure is nearly isotropic ($\beta\simeq 0$) inside the central $\rho\propto r^{-3/2}$ cusps.

We also explore the three-dimensional shapes of our haloes. We use the algorithm of \citet{2006MNRAS.367.1781A} \citep[see also][]{2011MNRAS.416.1377V}, which involves regarding a halo (or concentric subset thereof) as an ellipsoid and evaluating the associated axis lengths $l_1\geq l_2\geq l_3$ via the halo's inertia tensor. An ellipsoid's axis lengths are proportional to the square roots of the eigenvalues of its inertia tensor. Specifically, this procedure employs the ``reduced'' inertia tensor
\begin{equation}
    I_{ij} = \sum_{n}m_n \frac{x_{n}^{(i)}x_{n}^{(j)}}{\tilde r_n^2},
\end{equation}
summed over particles $n$ at ``ellipsoidal'' radii $\tilde r_n$, which avoids weighting more distant particles more highly. If $x_n$, $y_n$, and $z_n$ are coordinates in the ellipsoid's principal-axes frame, then $\tilde r_n^2\equiv (x_n/l_1)^2+(y_n/l_2)^2+(z_n/l_3)^2$. Beginning with a spherical ellipsoid with axes $l_1=l_2=l_3$, we evaluate $I_{ij}$ for particles inside the ellipsoid, evaluate the resulting new axis lengths (keeping the longest axis length $l_1$ fixed), and rotate the system into the frame of these new axes. This procedure is iterated until the axis ratios $l_2/l_1$ and $l_3/l_1$ are converged to better than $10^{-3}$.

The lower panels of Fig.~\ref{fig:aniso} show the axis ratios $l_3/l_1$ and $l_2/l_1$ as a function of the radius $r=(l_1 l_2 l_3)^{1/3}$, which is the radius of the sphere that encloses the same volume as the ellipsoid. Generally these haloes are highly aspherical, although there is a weak tendency for them to become more spherical over time, as measured by the $l_3/l_1$ axis ratio. At late times, these haloes tend to be less spherical at large radii and more spherical at small radii, a trend that is interestingly the opposite of what \citet{2011MNRAS.416.1377V} found for cold dark matter haloes. Also, our haloes tend to be comparatively prolate at large radii, as evidenced by the ratios $l_3/l_1$ and $l_2/l_1$ both lying well below unity. With respect to the $\rho\propto r^{-3/2}$ central cusps, however, there is little uniformity. H3's cusp is initially triaxial but becomes nearly spherical at late times. W5 and C3 have highly prolate cusps that remain so through their evolution, although the portion of W5's cusp at intermediate radii $r\gtrsim 10^{-2}$ becomes more oblate. W1's cusp is triaxial, although it appears to be evolving rapidly toward a more spherical shape.

Despite the variety of three-dimensional shapes of haloes' central cusps, we find one general feature: the cusp tends to align with the halo at larger radii. Figure~\ref{fig:shape} shows slices of the equidensity surfaces of the same four haloes at radii $r\sim 0.01$ and $r\sim 0.2$. While some of the ellipsoidal axis ratios are clearly more extreme at larger radii, the axes themselves are closely aligned between the inner cusp at $r\lesssim 0.01$ and the broader halo at $r\sim 0.1$.

\section{Minimum radius of the inner cusp}\label{sec:core}

We have found that the $\rho\propto r^{-3/2}$ density cusps of our haloes extend down to the resolution limits of our simulations, but it is unclear how much farther this power law could extend. If the dark matter were infinitely cold, but the initial power spectrum nevertheless had a small scale cutoff, then it is possible that the cusp could extend to arbitrarily small radii. However, any thermal motions of dark matter particles in the initial conditions would impose a minimum radius below which the $\rho\propto r^{-3/2}$ structure must give way to a core. We now explore what the radius of this core should be.

\subsection{Maximum phase-space density}\label{sec:f}

The thermal core can be understood as a consequence of Liouville's theorem, which implies that the maximum coarse-grained phase-space density cannot increase with time \citep{1979PhRvL..42..407T}. As a result, the (time-independent) maximum phase-space density $\fmax$ of the dark matter at times prior to formation of any non-linear structure sets an upper limit on the value of the distribution function $f(\vx,\vv)$ anywhere inside a halo at all later times \citep[e.g.][]{2001ApJ...561...35D,2006ApJ...652..306S,2011JCAP...03..024V,2012MNRAS.424.1105M}. Simulations suggest that halo centres approximately saturate this limit \citep{2012MNRAS.424.1105M}.

We first estimate $\fmax$. In general, dark matter that was once coupled to the radiation bath is left with some momentum distribution $f_p(\vp,a)$ that peaks at $\vp=0$ and has width proportional to some scale momentum $\pc=\pdc (\ad/a)$ that declines as $a^{-1}$ from its initial value $\pdc$, which was defined at the scale factor $\ad$ when the dark matter kinetically decoupled. The maximum value of this distribution is  given by $f_p(0)=C\pc^{-3}$, where $C$ is a normalization factor independent of $a$. Two specific examples are cold dark matter (CDM) and warm dark matter (WDM):
\begin{enumerate}
    \item In the CDM scenario, dark matter decouples while it is nonrelativistic. In this case it has a Maxwell-Boltzmann momentum distribution, $f_p(\vp) = (2\pi)^{-3/2} \pc^{-3}\exp[-\vp^2/(2\pc^2)]$, so that $C=(2\pi)^{-3/2}$ and $\pdc=(\Td m)^{1/2}$, where $\Td$ is the temperature at decoupling and $m$ is the particle mass.
    \item In the WDM scenario, dark matter is a fermion that decouples while it is ultrarelativistic. In this case it has an ultrarelativistic Fermi-Dirac momentum distribution, $f_p(\vp) = [6\pi\zeta(3)]^{-1} \pc^{-3}\left[\exp(p/\pc)+1\right]^{-1}$ (where $\zeta$ is the Riemann zeta function), so that $C=[12\pi\zeta(3)]^{-1}\simeq 0.0221$ and $\pdc=\Td$.
\end{enumerate}

Let us now consider a time when the dark matter is nonrelativistic (regardless of whether it was relativistic at decoupling). Its spatial density is then $\rhoM a^{-3}$, where $\rhoM$ is the present-day dark matter density (i.e. at $a=1$). Consequently, the maximum density in position-momentum phase space is $(C\pc^{-3})(\rhoM a^{-3}) = C\rhoM(\pdc\ad)^{-3}$. Within dark matter haloes we find it more convenient to work with the equivalent position-velocity phase-space density, the maximum value of which is therefore
\begin{equation}\label{fmax}
    \fmax = C\rhoM(\pdc\ad/m)^{-3},
\end{equation}
where $m$ is the mass of the dark matter particle.

We may now estimate a halo's thermal core radius $\rc$ as the largest radius at which its phase-space density reaches the value $\fmax$. If we also define $\rhoc\equiv \rho(\rc)$, then we show in Appendix~\ref{sec:structure} that
\begin{equation}\label{rhor6}
    \rhoc \rc^6\simeq 3\times 10^{-5} G^{-3}\fmax^{-2}
\end{equation}
for a halo with central cusp $\rho\propto r^{-\gamma}$ with $1\lesssim\gamma\lesssim 1.5$. Thus, once a halo's rescaled density profile $\rho r^6$ approaches $\rhoc\rc^6$, its central cusp must give way to a thermal core.

\subsection{Warm dark matter core sizes}\label{sec:coresim}

Let us now explore the core sizes that would develop in our simulated haloes if they were to form in a WDM cosmology. Eq.~(A3) of \citet{2001ApJ...556...93B} gives the characteristic nonrelativistic streaming velocity $(\pdc/m)(\ad/a)$ as a function of the present-day dark matter density $\Omega_\chi$ (in units of the critical density), the Hubble parameter $h$, the dark matter particle's effective number $n_\chi$ of degrees of freedom, and its mass $m$. We fix $m=3.5$~keV, $g_\chi=1.5$, $h=0.68$, and $\Omega_\chi=0.3$ (and for simplicity we assume that dark matter constitutes all of the matter). Consequently, $\pdc\ad/m=7.8\times 10^{-9}$ \chg{(taking the speed of light $c=1$)}, and hence by Eq.~(\ref{fmax}), $\fmax=1.8\times 10^{33}$~$\mathrm{M}_\odot$\,Mpc$^{-3}$. Now Eq.~(\ref{rhor6}) implies that the phase-space barrier lies at $\rhoc\rc^6=8.0\times 10^{-14}$~$\mathrm{M}_\odot$\,Mpc$^3$.

\begin{figure}
	\centering
	\includegraphics[width=\columnwidth]{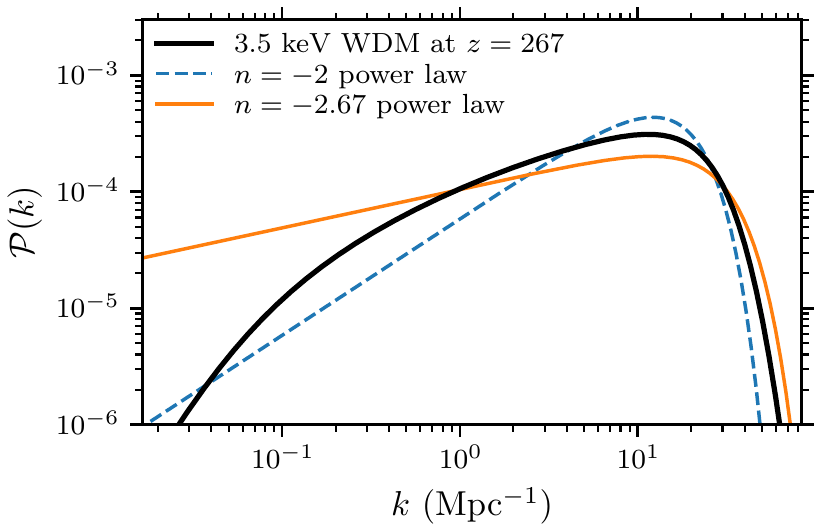}
	\caption{Matter power spectrum at redshift $z=267$ in a 3.5~keV WDM scenario (solid black curve) compared to appropriately rescaled versions of our $n=-2$ and $n=-2.67$ power-law simulation power spectra. In particular we rescale to match $\sigma$, the rms density contrast, and $\langle k^2\rangle$. The 3.5~keV WDM power spectrum appears to effectively lie somewhere between these two power laws.}
	\label{fig:wdm-power}
\end{figure}

The simulations, however, have been carried out with arbitrary units. To fix the units, we must compare our power spectra (Eq.~\ref{pk}) to the power spectrum that arises in the WDM scenario described above. We evaluate the matter power spectrum using the fitting form from \citet{1998ApJ...496..605E} and normalize it to $\sigma_8=0.81$ \citep{2020A&A...641A...6P} at $a=1$. Next, we use the transfer function described in \citet{2001ApJ...556...93B} to apply the small-scale cutoff corresponding to the WDM scenario above. In particular, we find that the free-streaming length $\alpha_\mathrm{fs}$ of \citet{2001ApJ...556...93B} in this scenario is $\alpha_\mathrm{fs}=0.017$~Mpc. We now recall that our simulations begin when the rms density contrast is $\sigma=0.03$. The WDM power spectrum described above must be scaled back to $z=267$ (using the growth function) to have the same rms density contrast. Consequently, we interpret the starting redshift of our simulations to be $z=267$, and their final redshifts range over $3.8<z<7.0$ for the $n=-2$ and $n=-2.67$ cases which are most relevant in this context. Figure~\ref{fig:wdm-power} shows the WDM power spectrum at the initial time (thick solid curve).

\begin{figure}
	\centering
	\includegraphics[width=\columnwidth]{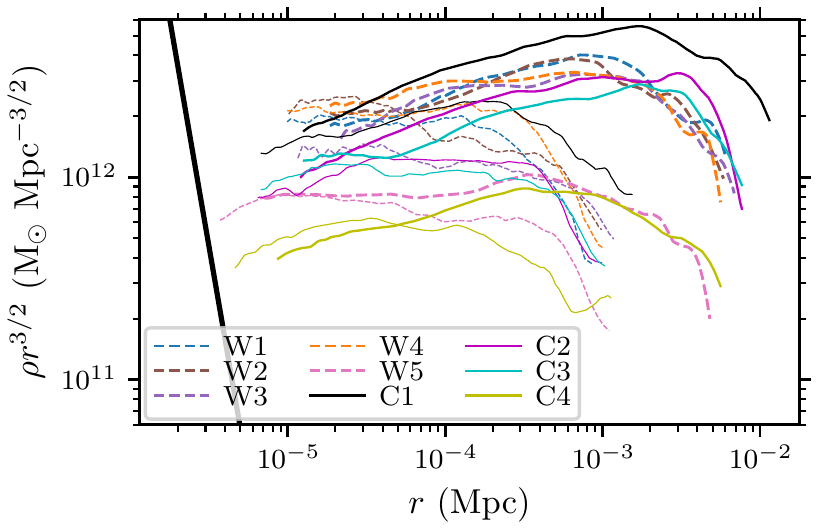}
	\caption{Predicted core sizes for our simulated haloes if they arise in a WDM scenario. In particular, we plot the density profiles $\rho(r)$ (scaled by $r^{3/2}$) for the haloes that arise from the $n=-2$ (dashed curves) and $n=-2.67$ (solid curves) power spectra. These haloes have been rescaled to correspond to a 3.5 keV WDM scenario (as described in the text). Each halo is plotted in a different colour, and for each halo we plot the density profile shortly after formation (thin curve) and a factor of about 1.9 in the scale factor later (thick curve). On the left, we show the $\rho r^6=\rhoc\rc^6$ line (black) corresponding to Eq.~(\ref{rhor6}). Each halo's inner cusp is expected to give way to a central core where it crosses this line.}
	\label{fig:cores-WDM}
\end{figure}

The remaining step is to fix the length scales in the simulations. Since the simulation power spectra have different forms than the WDM power spectrum, there is no unique way to make this association. We choose to fix $\langle k^2\rangle$ (Eq.~\ref{k2}; see the considerations in Section~\ref{sec:setup}), which evaluates to $\langle k^2\rangle=150$~Mpc$^{-2}$ for the WDM power spectrum. Consequently we take the simulation length unit to be $\langle k^2\rangle^{-1/2}=8.2\times 10^{-2}$~Mpc (see Table~\ref{tab:units}), which implies that the exponential cutoff wavenumber is $\kcut\simeq 17$~Mpc$^{-1}$ for $n=-2$ and $\kcut\simeq 29$~Mpc$^{-1}$ for $n=-2.67$. Figure~\ref{fig:wdm-power} shows the rescaled $n=-2$ (dashed curve) and $n=-2.67$ (thin solid curve) power spectra.

The WDM power spectrum evidently lies effectively somewhere between the $n=-2$ and $n=-2.67$ power spectra, so we consider the nine simulated haloes that arose therefrom. After applying the scaling described above, we plot density profiles for these haloes in Fig.~\ref{fig:cores-WDM}. We also plot the line $\rho r^6 = \rhoc\rc^6=8\times 10^{-14}$~$\mathrm{M}_\odot$\,Mpc$^3$; the interpretation is that each halo's inner cusp (as long as its slope lies between $\rho\propto r^{-3/2}$ and $\rho\propto r^{-1}$) must give way to a central core once it crosses this line. Evidently, each halo's $\rho\propto r^{-3/2}$ cusp should extend over about two decades in radius shortly after formation and up to almost three decades in radius after halo growth in those cases where the $r^{-3/2}$ cusp persists. In general, haloes in this WDM scenario have core radii of about 3~pc. At early times, we resolve the inner cusps of our haloes down to about three times the predicted core radius, although this resolution worsens as time goes on.

\subsection{Cold dark matter core sizes}

We also explore core sizes that would arise in a CDM cosmology. In particular, we suppose that the dark matter has particle mass $m=100$~GeV and kinetically decouples at the temperature $\Td=30$~MeV \chg{(so $\ad\simeq 5.3\times 10^{-12}$)}. In this case, $\pdc\ad/m=9.2\times 10^{-14}$ \chg{(again taking $c=1$)} and hence $\fmax=3.1\times 10^{48}$~$\mathrm{M}_\odot$\,Mpc$^{-3}$, which implies that the phase-space barrier lies at $\rhoc\rc^6=2.7\times 10^{-44}$~$\mathrm{M}_\odot$\,Mpc$^3$.

\begin{figure}
	\centering
	\includegraphics[width=\columnwidth]{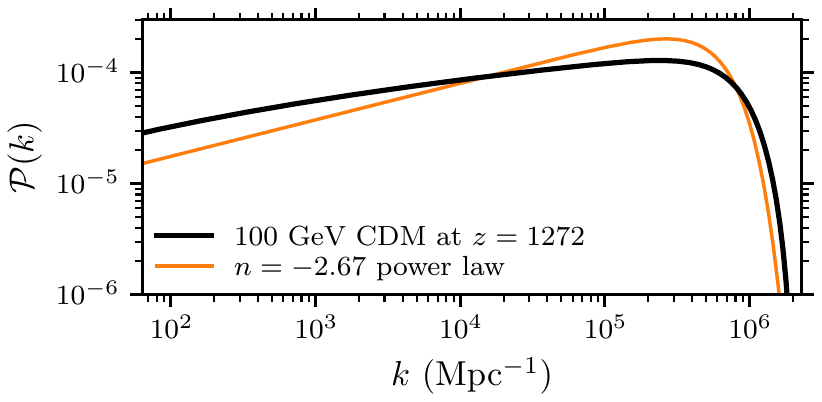}
	\caption{Matter power spectrum at redshift $z=1272$ in a 100~GeV CDM scenario (black) compared to a version of our $n=-2.67$ power-law simulation power spectrum that we rescale to match $\sigma$ and $\langle k^2\rangle$ (orange).}
	\label{fig:cdm-power}
\end{figure}

To find the power spectrum, we scale the aforementioned \citet{1998ApJ...496..605E} fitting function using the CDM transfer function in \citet{2005JCAP...08..003G}. \chg{We then set $\sigma=0.03$ by extrapolating this power spectrum back to $z=1272$ using the growth function that is valid in mixed matter-dark energy domination}.\footnote{At such early times, the radiation density also has a significant influence on the growth function, but since our simulations were carried out assuming matter domination, we neglect this effect in order to maintain consistency. Also, baryons do not cluster at the scale of the CDM cutoff, an effect that makes CDM density variations grow slightly more slowly on such scales \citep{1996ApJ...471..542H}. Since neither our simulations nor the \citet{1998ApJ...496..605E} power spectrum that we use include this effect, we also neglect it in the growth function.} The resulting power spectrum is shown as the black curve in Fig.~\ref{fig:cdm-power}. This power spectrum has $\langle k^2\rangle=7.4\times 10^{10}$~Mpc$^{-2}$, so we take our simulation length unit to be $3.7\times 10^{-6}$~Mpc.

\begin{figure}
	\centering
	\includegraphics[width=\columnwidth]{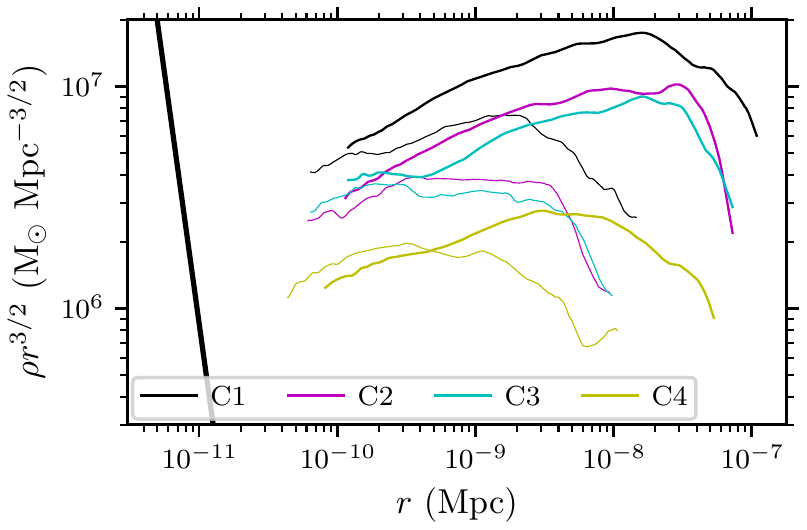}
	\caption{Predicted core sizes for our simulated haloes if they arise in a CDM scenario. We plot the density profiles $\rho(r)$ (scaled by $r^{3/2}$) for the haloes that arise from the $n=-2.67$ power spectrum, and these haloes have been rescaled to correspond to a 100 GeV CDM scenario with kinetic decoupling at 30 MeV. As in Fig.~\ref{fig:cores-WDM}, each halo is represented by a different colour, and for each halo we plot the density profile shortly after formation (thin curve) and a factor of about 1.9 in the scale factor later (thick curve). We show on the left the $\rho r^6=\rhoc\rc^6$ barrier (black) at which each halo's inner cusp is expected to give way to a central core.}
	\label{fig:cores-CDM}
\end{figure}

Figure~\ref{fig:cdm-power} shows the appropriately rescaled version of our $n=-2.67$ power spectrum (orange curve). This power spectrum is somewhat too steep to faithfully represent the CDM scenario, but the mismatch is not serious. We now show in Fig.~\ref{fig:cores-CDM} the appropriately rescaled density profiles of the haloes arising from the $n=-2.67$ power spectrum along with the corresponding phase-space barrier $\rho r^6=2.7\times 10^{-44}$~$\mathrm{M}_\odot$\,Mpc$^3$. Core radii in this scenario evidently lie around $10^{-5}$~pc, where the profiles would cross the phase-space barrier. Interestingly, despite the huge difference in scales between the first haloes in these WDM and CDM scenarios, the core radii are very similar when considered as fractions of the halo (or cusp) radii. CDM core radii are only about three times smaller than WDM cores in this relative sense. Like in the WDM case, the $\rho\propto r^{-3/2}$ cusps of CDM haloes can extend across two to three decades in radius in cases where they persist.

\subsection{Scaling behaviour}

The observation that CDM and WDM core sizes are nearly the same, when considered in relation to the extent of the halo's inner cusp, merits further discussion. In a general sense, it is expected that core sizes and cusp sizes scale together because the former are set by the dark matter thermal velocity distribution while the latter are set by the drift distances that result from these velocities. However, it is not obvious that their mutual scaling should be so tight.

Consider dark matter particles with mass $m$ that kinetically decoupled at the scale factor $\ad$ during the last radiation-dominated epoch with characteristic (possibly relativistic) momentum $\pdc$. Their characteristic comoving drift distance is
\begin{equation}\label{lambda}
    \lambda=\int_{\ad}^a \diff a^\prime
    \frac{\left[1+\left(\frac{\pdc\ad}{m a^\prime}\right)^{-2}\right]^{-1/2}}{a^{\prime 2} H(a^\prime)}
    =
    \left(\frac{8\pi G}{3}\rhoM\aeq\right)^{-1/2}\frac{\pdc\ad}{m}\Lfs
\end{equation}
by the scale factor $a$, where
\begin{equation}\label{Lfs}
    \Lfs \simeq \ln\left[\frac{8}{1\!+\!\sqrt{1\!+\!(\pdc/m)^2}}\frac{\aeq}{\ad}\right]
\end{equation}
is roughly the logarithm of the factor in $a$ over which particles stream nonrelativistically during radiation domination. Here $H(a)$ is the Hubble rate, $\rhoM$ is the matter density at $a=1$ today, and $\aeq$ is the scale factor at matter-radiation equality. In Eq.~(\ref{Lfs}) we assume the dark matter becomes nonrelativistic long before $\aeq$ and that we are evaluating $\lambda$ at a scale factor $a\gg \aeq$. We also neglect changes in the radiation content, which in general raise $\Lfs$ by a small amount.\footnote{$H$ is scaled by $(g_*/3.36)^{1/2}(g_{*s}/3.91)^{-2/3}$, where $g_*$ and $g_{*s}$ are the effective numbers of relativistic degrees of freedom for energy and entropy density, respectively. Assuming only Standard Model content, this factor is $1$ at temperatures below about 1~MeV, roughly 0.9 at temperatures between 1~MeV and 100~MeV, and roughly 0.6 at temperatures above about 100~MeV \citep[e.g.][]{2016Galax...4...78H}. Since $\lambda\sim H^{-1}$, inclusion of this effect can boost $\lambda$ by up to a factor of about 2, although typically the effect is far smaller because $H^{-1}$ appears inside an integral. Warm dark matter, however, can require large $g_*$ at decoupling to give dark matter the observed relic abundance.}

If $R$ is the characteristic (comoving) size of a peak in the primordial density field smoothed by free streaming, we expect that $R\sim\lambda$. For a halo that collapses at the scale factor $\ac$ from such a peak, we thus expect that the halo's $\rho\propto r^{-3/2}$ cusp has coefficient $A=\rho r^{3/2}\sim \rhoM \ac^{-3/2} \lambda^{3/2}$ and initial radius $r_0\sim \ac\lambda$ (see Section~\ref{sec:peaks}). Now through these relations and Eqs. (\ref{fmax}), (\ref{rhor6}), and~(\ref{lambda}) we find that
\begin{align}\label{corecusp}
    \rc/r_0 \sim (\ac/\aeq)^{-2/3}\Lfs^{-4/3}.
\end{align}
The mutual scaling between core radius $\rc$ and cusp radius $r_0$ thus emerges, and we see that there are two main influences that can raise $r_0$ without raising $\rc$.\footnote{$\rc/r_0$ is also sensitive to the shape of the initial momentum distribution, both through the normalization factor $C$ (see Section~\ref{sec:f}) and the form of the power spectrum cutoff, which sets the connection between $\lambda$ and the distribution of peak radii $R$. The $R$ distribution also depends on the underlying initial power spectrum of density variations. These are order-unity considerations, however.} The first possibility is to raise $\Lfs$ by increasing the duration of nonrelativistic drift (or by raising the number of relativistic degrees of freedom in the early universe). This change raises the free-streaming length without altering particle velocities and hence $\fmax$. The second possibility is to form haloes later, which raises the physical free-streaming scale $\ac\lambda$ without altering $\fmax$.

CDM and WDM have comparable values of $\Lfs$ (of order 10) and $\ac/\aeq$ (of order 100), so they approximately preserve the mutual scaling between $\rc$ and $r_0$. However, nonstandard early-universe physics can break this scaling to a more serious extent. For instance, certain inflationary \citep[e.g.][]{2010JCAP...09..007B} or postinflationary \citep[e.g.][]{2011PhRvD..84h3503E} scenarios can greatly boost the power spectrum at small scales, potentially leading to $\ac\sim\aeq$ for the first haloes, \chg{or even $\ac<\aeq$ \citep[e.g.][]{2019PhRvD.100j3010B,2022arXiv221004904D}}. Moreover, if the dark matter decoupled in the presence of a large number of beyond-Standard Model relativistic degrees of freedom, this effect could significantly raise $\Lfs$. A large injection of entropy after dark matter decoupling, such as that arising from a period of early matter domination \citep[e.g.][]{2021OJAp....4E...1A}, has a similar effect. Also, these considerations only apply to haloes that form at the free-streaming scale. Inflationary dynamics \citep[e.g.][]{2009PhRvD..80l6018B} can give rise to features in the primordial power spectrum that are completely independent of the free-streaming scale, and similar features can arise in the matter power spectrum due to postinflationary physics \citep[e.g.][]{1988PhLB..205..228H,2021PhRvD.103j3508E}. If such a feature is sufficiently pronounced to form haloes by direct collapse, those haloes could develop $\rho\propto r^{-3/2}$ density profiles with arbitrarily small cores.

\section{Conclusions}\label{sec:conclusion}

We have carried out high-resolution cosmological zoom simulations of twelve first-generation haloes of mass corresponding approximately to the small-scale cutoff in the fluctuation power spectrum. The three idealized linear power spectra we considered correspond roughly (but not exactly) to those associated with hot, warm, and cold dark matter. Our main findings are the following:
\begin{enumerate}
    \item Soon after collapse and for all three initial power spectra, almost all our haloes contained central cusps with $\rho\propto r^{-3/2}$ and with total mass comparable to that of the peak in the linear density field from which the halo originated. We could follow this cusp over as many as two decades in radius. Cusp formation is independent both of collapse morphology and of the prominence of small-scale N-body clumping artefacts in the pre-collapse structure. Collapses through a filament, through a sheet, and through more complex caustic networks all produce similar cusps. We explicitly verified in one case that the cusp profile is stable with respect to changes in the simulation resolution.
    \item Since the cusp forms so quickly after collapse, its coefficient $A=\rho r^{3/2}$ is sensitive almost wholly to the immediate neighborhood of the halo's precursor density peak. This outcome confirms earlier findings by \citet{2019PhRvD.100b3523D}. In particular, we find the peak's height and spherically averaged characteristic radius, along with the tidal tensor at the location of the peak, suffice to determine $A$ to within 10 to 20 per cent.
    \item Haloes that initially possess $\rho\propto r^{-3/2}$ central cusps tend to develop shallower density profiles as they grow. However, unlike previous works that attributed this process to major mergers, we find that it can be attributed to rapid accretion in general (which includes major mergers, minor mergers, and accretion of the smooth background). In fact, this shallowing is consistent with simple models of the sort presented by \citet{2010arXiv1010.2539D} and \citet{2013MNRAS.432.1103L} that connect halo density profiles to their accretion histories. But unlike mergers, rapid accretion tends to build up a halo's density profile at intermediate and large radii without disturbing its inner profile; it does not necessarily disrupt the initial $\rho\propto r^{-3/2}$ cusp.
    \item Major mergers can disturb the central $\rho\propto r^{-3/2}$ cusp, as \citet[][]{2016MNRAS.461.3385O} and \citet{2017MNRAS.471.4687A} found. In two cases, we found that a major merger event reduced the amplitude of the central $\rho\propto r^{-3/2}$ cusp by about 10 per cent (although we do not have the resolution to assess whether the cusp's slope changed). In the absence of mergers, on the other hand, the central cusp remains undisturbed even when accretion is otherwise rapid. We also found a case where even a major merger had no impact on the central cusp. Within our halo sample and resolution limits, the extent to which mergers disrupt the $\rho\propto r^{-3/2}$ cusp remains small.
    \item For haloes that grow sufficiently slowly, the $\rho\propto r^{-3/2}$ central cusp can continue to comprise between 1 and 10 per cent of the total halo mass even long after the halo's formation. Two of the haloes arising from the hot dark matter-like power spectrum maintained a cusp mass fraction in this range over a factor of about 10 in the scale factor. On the other hand, a period of rapid accretion tends to swiftly reduce the cusp mass fraction below $10^{-3}$.
    \item The $\rho\propto r^{-3/2}$ cusp invariably has a nearly isotropic velocity dispersion. Conversely, it exhibits a wide range of three-dimensional shapes across different haloes and even for the same halo across different times. Equidensity surfaces within the cusp tend to align with those of the halo at larger radii.
    \item It is still unclear how far down in radius the $\rho\propto r^{-3/2}$ cusp can extend, owing to the limits of our simulation resolution. However, if the dark matter was once in contact with the radiation bath, then its residual thermal motion means that any density cusp must give way to a finite-density core at small enough radii. Some of our simulations probe radii down to as little as three times the expected core radius in a warm dark matter scenario, which suggests that the question of how much farther the $\rho\propto r^{-3/2}$ cusp can extend may be largely immaterial, at least for conventional dark matter models in which the power spectrum's cutoff arises from thermal free streaming. Nevertheless, the core radius is small enough that the $\rho\propto r^{-3/2}$ density profile can be important if it survives.
\end{enumerate}

Survival of the $\rho\propto r^{-3/2}$ central cusp through periods of rapid accretion can lead to density profiles resembling that in Fig.~\ref{fig:W1evo}. There, a halo with an Einasto density profile at intermediate and large radii still possesses its natal $\rho\propto r^{-3/2}$ cusp at the smallest radii. In larger-scale cosmological simulations, this cusp might go unresolved, which could explain why some works \citep[e.g.][]{2020Natur.585...39W} found no evidence of $\rho\propto r^{-3/2}$ cusps despite resolving the smallest haloes. Moreover, this picture demonstrates a new avenue by which the $\rho\propto r^{-3/2}$ cusps of the first haloes may transition into the Einasto density profiles seen at larger scales. Since this mechanism does not destroy the cusp, this discovery raises the prospect that a significant number of $\rho\propto r^{-3/2}$ cusps could survive up to the present day.

We studied only a small halo sample over a limited time range. More work is needed to determine the true extent to which the $\rho\propto r^{-3/2}$ cusps of the first haloes persist through cosmic time. If these cusps survive, they could have observational implications. The abundance of the smallest subhaloes could be larger than previously estimated, owing to the lower efficiency of tidal stripping for steep central cusps \citep{2010MNRAS.406.1290P,2022arXiv220700604S}. For instance, \citet{2021JHEP...06..028L} concluded that the suppression of the smallest CDM subhaloes by tidal stripping pushes them outside the reach of pulsar timing searches. This conclusion may need to be reassessed if $\rho\propto r^{-3/2}$ density profiles persist. Separately from the question of subhalo survival, observable signatures of dark matter haloes, such as through dark matter annihilation \citep[e.g.][]{2010ApJ...723L.195I,2022arXiv220911237D} and astrometric microlensing \citep{2011ApJ...729...49E}, tend to be larger if they have steeper cusps. We also comment that these considerations are amplified for nonstandard cosmological initial condition scenarios in which small-scale density variations are boosted \citep[e.g.][]{2011PhRvD..84h3503E}. The haloes that arise at those boosted scales grow slowly, so their $\rho\propto r^{-3/2}$ cusps may remain a large fraction of the total mass in haloes.

Finally, we have not addressed the question of why the $\rho\propto r^{-3/2}$ density profile arises. This profile appears to be a general consequence of the collapse of a smooth density peak, independent of the details of the collapse process. It has been observed that the Einasto or NFW density profiles of idealized CDM haloes can largely be explained by appealing to their accretion histories \citep[][]{2010arXiv1010.2539D,2013MNRAS.432.1103L}, an idea that is linked to the notion that material stratifies within the halo based on its accretion time (see Section~\ref{sec:accretion}). However, this idea cannot explain the $\rho\propto r^{-3/2}$ density profiles of the first haloes; these profiles form immediately upon collapse, before there is any meaningful accretion history. A new idea is needed \citep[e.g.][]{2022MNRAS.517L..46W}.

\section*{Acknowledgements}

We thank Neal Dalal, Benedikt Diemer, and Go Ogiya for useful discussions. The simulations for this work were carried out on the MPA's Freya computing cluster at the Max Planck Computing and Data Facility.

\section*{Data Availability}
 
Simulation data are available upon reasonable request to the corresponding author.



\bibliographystyle{mnras}
\bibliography{main}



\appendix

\section{Mixing of low- and high-resolution particles}\label{sec:lowres}

\begin{figure}
	\centering
	\includegraphics[width=\columnwidth]{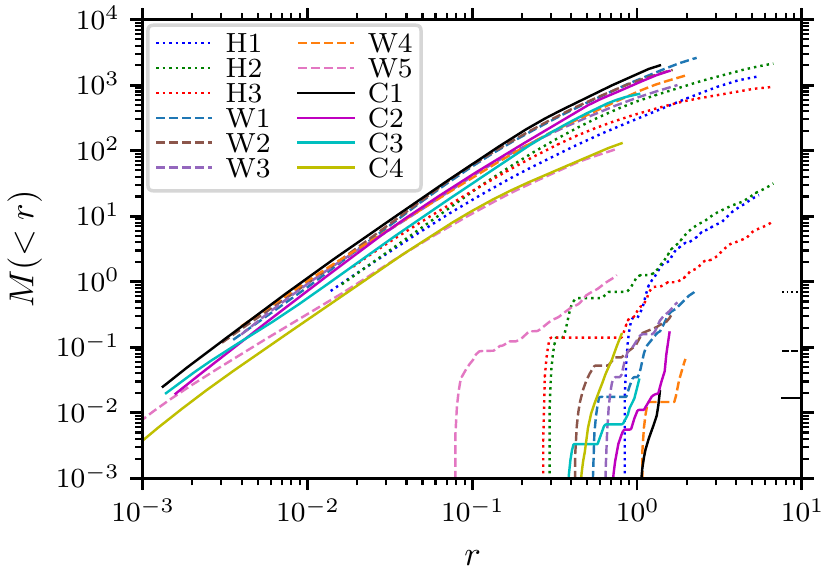}
	\caption{Total enclosed mass profiles of our haloes (diagonal curves through the centre) along with the mass profiles contributed by low-resolution particles only (lower right). We average these profiles over the final factor of 1.17 in the scale factor (i.e. from $a_\mathrm{f}/1.17$ to $a_\mathrm{f}$). Black tick marks in the lower right show for each primary (low-resolution) simulation the mass of one low-resolution particle (but since we time average it, the mass profile of low-resolution particles is not required to advance in integer multiples thereof). Evidently, low-resolution particles comprise $\lesssim 1$ per cent of each halo, and all such particles reside at radii too large to affect this study's conclusions (compare Fig.~\ref{fig:all_late}).}
	\label{fig:particletype}
\end{figure}

Section~\ref{sec:setup} describes how we prepare our high-resolution halo simulations. After choosing the halo in a low-resolution simulation, we identify the Lagrangian region spanned by halo particles and replace all of the low-resolution particles in that region with much more numerous high-resolution particles. To spare computational expense, however, we do not make the high-resolution region so large as to ensure that low-resolution particles never fall into the high-resolution halo. In this appendix, we show that any numerical artefacts that arise from this choice cannot impact our results.

Mixing of particles of different masses can accelerate two-body relaxation effects \citep[e.g.][]{2008gady.book.....B}. Artificial energy exchange between simulation particles causes the more massive particles to gradually sink to the centre of the system while the light particles are dynamically heated. However, we plot in Fig.~\ref{fig:particletype} the total enclosed mass profile $M(r)$ of each of our haloes along with the mass profile contributed by low-resolution particles at the final scale factor $a_\mathrm{f}$. Evidently, each halo's low-resolution particles contribute 1 per cent or less of the total mass. Moreover, all such particles reside at radii so large that this small mass fraction cannot alter the conclusions of our study. Comparison with Fig.~\ref{fig:all_late} shows that, for all haloes, the low-resolution particles lie outside both the central $\rho\propto r^{-3/2}$ cusp and any shallower portion of the density profile created by rapid accretion.

\section{Numerical convergence at small radii}\label{sec:convergence}

\begin{figure*}
	\centering
	\includegraphics[width=\linewidth]{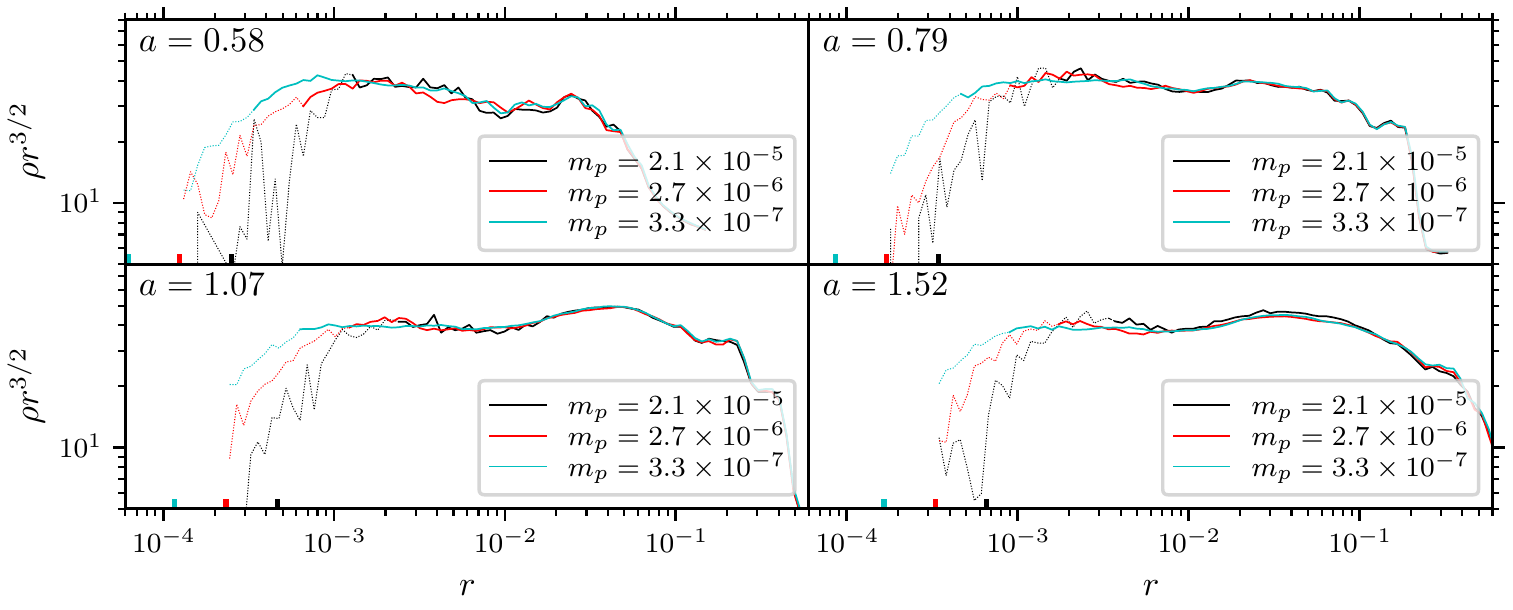}
	\caption{Density profile of the W5 halo at four different times (panels) and three different simulation resolutions (colours), indicated by the particle mass $m_p$. The force-softening length $\epsilon$ is scaled as $m_p^{1/3}$. While the density profiles all transition into a uniform-density core at small radii (indicated by a steeply rising curve in this scaled plot), this transition is evidently artificial because the transition radius scales with the simulation resolution. Meanwhile, the $\rho\propto r^{-3/2}$ behaviour (horizontal) is stable with respect to simulation resolution. We plot the density profile at radii below $5\epsilon$ with a thin dotted line; $\epsilon$ itself is marked for each simulation by a coloured tick at the bottom of each panel. We also plot radii above $5\epsilon$ but below the relaxation limit $r_\mathrm{rel}$ (where the relaxation time scale becomes shorter than the age of the universe; see the text) with a thin dashed line, although for this halo the only example occurs for $a=1.52$ (lower right-hand panel) at the lowest simulation resolution (black).}
	\label{fig:W5conv}
\end{figure*}

\begin{figure*}
	\centering
	\includegraphics[width=\linewidth]{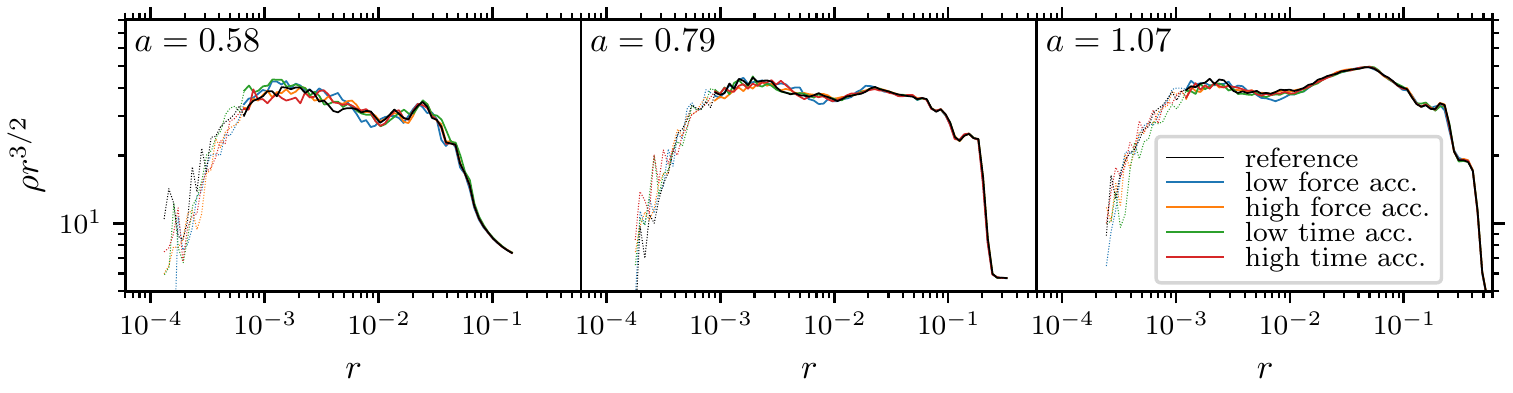}
	\caption{\chg{Density profile of the W5 halo at three different times (panels) and a variety of different \textsc{Gadget-4} numerical parameters (colours). We vary the tree-force accuracy and time integration accuracy as described in Appendix~\ref{sec:convergence}, but the impact on the density profile is marginal at early times (left-hand panel) and even smaller at later times (other panels).}}
	\label{fig:W5conv2}
\end{figure*}

We simulated halo W5 at the highest spatial resolution of all of our haloes (see Table~\ref{tab:halos}). In this section we test numerical convergence by simulating the same halo at worse resolution; specifically we raise the simulation particle mass $m_p$ and the force softening length $\epsilon$. We carried out two simulations in which we raise the particle mass by factors of 8 and 64 and the softening length by factors of 2 and 4, respectively. Figure~\ref{fig:W5conv} plots this halo's density profile at four different times in the original simulation (cyan) and in both reduced-resolution simulations (red and black). All density profiles transition into a core at small radii, but this core is clearly artificial in that its radius scales with the spatial resolution. The $\rho\propto r^{-3/2}$ part of the density profile, on the other hand, appears to be converged with respect to simulation resolution at all four times.

We find that the radius above which the density profile is converged is reasonably well approximated by $5\epsilon$, where $\epsilon$ is the force-softening length. Figure~\ref{fig:W5conv} depicts the density profile at radii below $5\epsilon$ with a thin dotted line. The idea here is that force softening introduces a bias \citep[e.g.][]{2001MNRAS.324..273D} that makes systems less tightly bound, and the relative effect becomes larger as we consider systems closer to the softening scale. However, this criterion somewhat fails at the earliest times for reasons that are unclear. In particular, the solid red and cyan curves in the upper left-hand panel begin to shallow at small radii $r$ even when $r>5\epsilon$, but this shallowing occurs at different radii for different resolution levels. Consequently, tendencies of some of our haloes' early density profiles to shallow at the smallest radii, e.g. in Fig.~\ref{fig:profiles-coll}, should be interpreted with some skepticism.

Also, while the $5\epsilon$ condition mostly suffices for W5, simulations carried out over a longer duration must also consider two-body relaxation between simulation particles. We evaluate the relaxation time scale
\begin{equation}\label{trel}
    t_\mathrm{rel}(r)\simeq\frac{N}{8\log(r/\epsilon)} \frac{2\pi r}{\sqrt{G M/r}}
\end{equation}
\citep[e.g.][]{2008gady.book.....B,2003MNRAS.338...14P} as a function of radius $r$, where $M$ and $N$ are the mass and particle count enclosed within $r$. The radius $r_\mathrm{rel}$ at which $t_\mathrm{rel}$ is equal to the age of the universe (a decent approximation for the age of any halo) then serves as an alternative minimum converged radius. In Fig.~\ref{fig:W5conv}, radii with $5\epsilon<r<r_\mathrm{rel}$ are shown in principle with a thin dashed line, but only at $a=1.52$ (bottom right) for the lowest-resolution simulation (black) does an example of such a radius appear. This restriction becomes more relevant for haloes simulated for longer durations (see Fig.~\ref{fig:halomass}).

\chg{Since standard choices of numerical accuracy parameters are not widely validated in simulations of the first haloes, we also briefly explore the impact of \textsc{Gadget-4}'s time stepping and short-range force accuracy parameters \citep[see also][]{2018MNRAS.473.4339O,2018PhRvD..98f3527D}. For the simulations in the text, we set the tree-force accuracy parameter to be $\alpha_\mathrm{f}=0.005$, where tree nodes are only opened if the estimated fractional force error from truncating the calculation is more than $\alpha_\mathrm{f}$. We also set the adaptive time stepping parameter to be $\eta_\mathrm{t}=0.025$, which roughly ensures that a particle's acceleration induces a displacement smaller than $\eta_\mathrm{t}\epsilon$ over one time step, and we set the maximum time step to be $[\Delta\log a]_\mathrm{max} = 0.01$. In Fig.~\ref{fig:W5conv2}, we separately test $\alpha_\mathrm{f}=0.015$ (low force accuracy); $\alpha_\mathrm{f}=0.002$ (high force accuracy); $\eta_\mathrm{t}=0.075$ and $[\Delta\log a]_\mathrm{max} = 0.03$ (low time integration accuracy); and $\eta_\mathrm{t}=0.01$ and $[\Delta\log a]_\mathrm{max} = 0.004$ (high time integration accuracy). Evidently, the impact of these changes on the density profile of halo W5 is marginal at early times and essentially negligible at late times.}

\chg{Finally, we remark that density cusps of the first haloes at least as steep as $\rho\propto r^{-3/2}$ have been confirmed to arise not only in $N$-body simulations but also in simulations that solve the Vlasov-Poisson equations directly \citep{2021A&A...647A..66C}. Consequently, they cannot be entirely an $N$-body artefact.}

\section{Phase-space structure of a density cusp}\label{sec:structure}

In this appendix, we quantify the phase-space structure of a density cusp for the purpose of determining the size of the core that arises from phase-space conservation. Consider an equilibrium spherical dark matter system with the density profile
\begin{equation}\label{rhocusp}
\rho(r) = A r^{-\gamma}.
\end{equation}
The coefficient $A$ is the system's sole scaling parameter with dimensions of $\text{density}\times\text{length}^\gamma$. If $\gamma<3$, then the system's enclosed mass profile is
\begin{equation}\label{Mcusp}
    M(r)=\frac{4\pi}{3-\gamma}A r^{3-\gamma}.
\end{equation}
If $\gamma<2$, then the system's potential is
\begin{align}\label{Phicusp}
    \Phi(r)=\frac{4\pi}{(3-\gamma)(2-\gamma)} G A r^{2-\gamma},
\end{align}
and dimensional considerations imply that the velocity dispersion has the same scaling, $\sigma^2\propto r^{2-\gamma}$. Consequently, the distribution function scales as $f\propto \rho/\sigma^3 \propto r^{\gamma/2-3}\propto \Phi^{-1/2-2/(2-\gamma)}$. Let us assume that the distribution function is isotropic, so it depends on energy $E=\Phi+v^2/2$ alone, i.e. $f(r,v)\propto E^{-1/2-2/(2-\gamma)}$. Recall from Section~\ref{sec:3d} that the velocity dispersion inside the central cusp is indeed nearly isotropic. If we also assume that $\gamma>0$, then we can normalize $f$ by requiring that $\int\diff^3 \vv f(r,v) = \rho(r)$. Then
\begin{align}\label{f_general}
    f(r,v) 
    &=
    (2\pi)^{-3/2}
    \frac{
    \Gamma\left(\frac{\gamma}{2-\gamma}\!+\!\frac{3}{2}\right)
    }{
    \Gamma\left(\frac{\gamma}{2-\gamma}\right)
    }
    \frac{\rho(r)}{\Phi(r)^{3/2}}
    \left[1+\frac{v^2}{2\Phi(r)}\right]^{-\frac{1}{2}-\frac{2}{2-\gamma}}\!,
\end{align}
where $\Gamma$ is the gamma function.

\begin{figure}
	\centering
	\includegraphics[width=\columnwidth]{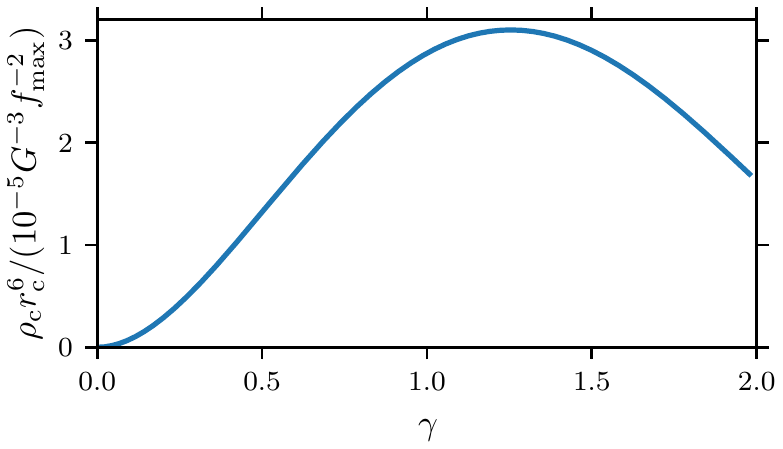}
	\caption{Sensitivity of the phase-space barrier $\rhoc \rc^6$ to the density profile slope $\gamma$. In particular, the maximum phase-space density within a $\rho\propto r^{-\gamma}$ density cusp begins to exceed $\fmax$ where $\rho r^6$ drops below $\rhoc\rc^6$. For the typical slopes $1\lesssim\gamma\lesssim 1.5$ observed in this work and others, $\rhoc\rc^6\simeq 3\times 10^{-5} G^{-3}\fmax^{-2}$.}
	\label{fig:corecoef}
\end{figure}

At each radius $r$, the maximum phase-space density is obtained by setting $v=0$:\footnote{Interestingly, Eq.~(\ref{fmax_general}) implies that spatially denser haloes (with larger $A$) have lower phase-space density, owing to their higher velocity dispersion.}
\begin{align}\label{fmax_general}
    f(r,0) 
    =
    \frac{\left[(3\!-\!\gamma)(2\!-\!\gamma)\right]^{3/2}}{2^{9/2}\pi^3}
    \frac{
    \Gamma\left(\frac{\gamma}{2-\gamma}\!+\!\frac{3}{2}\right)
    }{
    \Gamma\left(\frac{\gamma}{2-\gamma}\right)
    }
    G^{-3/2} A^{-1/2} r^{\gamma/2-3}
\end{align}
(where we have substituted Eqs. \ref{rhocusp} and~\ref{Phicusp}). If some maximum phase-space density $\fmax$ is set by dark matter microphysics, as discussed in Section~\ref{sec:core}, then the radius $\rc$ at which $\fmax$ is reached is thus
\begin{align}\label{rc_general}
    \rc 
    = 
    \left[\frac{(3\!-\!\gamma)(2\!-\!\gamma)}{8\pi^2}\right]^{\frac{3}{6-\gamma}}
    \left[\frac{
    \Gamma\left(\frac{\gamma}{2-\gamma}\!+\!\frac{3}{2}\right)
    }{
    \Gamma\left(\frac{\gamma}{2-\gamma}\right)
    }\right]^{\frac{2}{6-\gamma}}
    (G^3 A \fmax^2)^{-\frac{1}{6-\gamma}}.
\end{align}
If we define $\rhoc=\rho(\rc)$ and substitute $A=\rhoc\rc^{\gamma}$, then
\begin{equation}\label{rhor6_general}
    \rhoc \rc^6 = 
    \left[\frac{(3-\gamma)(2-\gamma)}{8\pi^2}\right]^3
    \left[\frac{
    \Gamma\left(\frac{\gamma}{2-\gamma}\!+\!\frac{3}{2}\right)
    }{
    \Gamma\left(\frac{\gamma}{2-\gamma}\right)
    }\right]^2
    G^{-3}\fmax^{-2}
\end{equation}
defines the general barrier in $\rho$--$r$ space at which $\fmax$ is reached. The right-hand side of this equation depends only on $\fmax$ and not on the parameters of the particular halo, other than $\gamma$. In fact, within the range $1<\gamma<1.5$, $\rhoc \rc^6$ is almost independent of $\gamma$ as well. We plot its dependence on $\gamma$ in Fig.~\ref{fig:corecoef}; evidently,
\begin{equation}\label{rhor6_approx}
    \rhoc \rc^6\simeq 3\times 10^{-5} G^{-3}\fmax^{-2}
\end{equation}
for any halo within this slope range. 

\label{lastpage}
\end{document}